\documentclass[twocolumn]{aastex631}

\usepackage{amsmath}
\usepackage{multirow}
\usepackage{xcolor} 
\usepackage{hyperref}

\usepackage{tikz}
\usetikzlibrary{arrows.meta,positioning,fit,calc,backgrounds}

\begin{document}

\title{Beyond what's observed: hierarchical inference of Gaia astrometric compact-object binaries}

\author[0000-0001-6970-1014]{Natsuko Yamaguchi}
\affiliation{Department of Astronomy, California Institute of Technology, 1200 E. California Blvd, Pasadena, CA, 91125, USA}
\email{nyamaguc@caltech.edu}

\author[0000-0002-6871-1752]{Kareem El-Badry}
\affiliation{Department of Astronomy, California Institute of Technology, 1200 E. California Blvd, Pasadena, CA, 91125, USA}

\author[0000-0001-8196-9267]{Reed Essick}
\affiliation{Canadian Institute for Theoretical Astrophysics, University of Toronto, 60 St. George Street, Toronto, ON M5S 3H8}
\affiliation{Department of Physics, University of Toronto, 60 St. George Street, Toronto, ON M5S 1A7}
\affiliation{David A. Dunlap Department of Astronomy, University of Toronto, 50 St. George Street, Toronto, ON M5S 3H4}

\author[0000-0002-6121-0285]{Amanda Farah}
\affiliation{Canadian Institute for Theoretical Astrophysics,
University of Toronto, 60 St. George Street, Toronto, ON M5S 3H8}

\author[0000-0003-1247-9349]{Cheyanne Shariat}
\affiliation{Department of Astronomy, California Institute of Technology, 1200 E. California Blvd, Pasadena, CA, 91125, USA}

\author[0000-0001-9298-8068]{Sahar Shahaf}
\affiliation{Max-Planck Institute for Astronomy, Königstuhl 17, D-69117 Heidelberg, Germany}

\begin{abstract}
Astrometry from the Gaia mission has revealed a large and homogeneous population of binaries containing compact objects, including more than 3000 white dwarf--main sequence (WDMS) binaries with orbital periods of $P_{\rm orb}\sim100$--$1000$ d. Most of these binaries likely underwent mass transfer from asymptotic giant branch donors, but their evolutionary histories remain uncertain. We develop a hierarchical Bayesian inference model to constrain the intrinsic properties of the Gaia WDMS binary population while accounting for measurement uncertainties and the survey selection function. We infer the distributions of orbital period, component masses, and eccentricity, as well as the total space density. We find that the intrinsic orbital-period distribution follows a declining power law, $p(P_{\rm orb})\propto P_{\rm orb}^{-0.445\pm0.048}$, and the WD mass distribution is sharply peaked near $0.6\,M_\odot$, with only $\sim1\%$ of systems hosting WDs more massive than $0.8\,M_\odot$. The inferred MS-star mass distribution is relatively flat down to $\sim0.3\,M_{\odot}$, which is unexpected if the population is dominated by products of stable mass transfer with a fixed critical mass ratio.  The eccentricity distribution peaks at a nonzero value, $e=0.046\pm0.001$, suggesting incomplete circularization or eccentricity excitation. We infer a midplane space density of $\sim3\times10^4\,{\rm kpc}^{-3}$ for WDMS binaries in the parameter space probed by Gaia DR3. These properties provide empirical benchmarks for future work to constrain binary evolution models. We also apply the framework to Gaia black hole and neutron star binaries, inferring that $30^{+66}_{-20}$ and $614^{+203}_{-149}$ such systems exist within $2\,$kpc, respectively. This framework will enable detailed population-level constraints on a wide range of systems identified from the much larger catalog of astrometric orbits expected from Gaia DR4. 
\end{abstract}

\keywords{Binary stars (154) --- White dwarf stars (1799) --- Astrometry (80)}

\section{Introduction} \label{sec:intro}

\defcitealias{Shahaf2024MNRAS}{S24}  
\defcitealias{Yamaguchi2026arXiv}{Y26} 

The third data release (DR3) of the Gaia mission included astrometric orbits for more than 160,000 binaries \citep{GaiaCollaboration2023A&A}, representing a dramatic increase in sample size over all previous work. This rich dataset has led to the discovery of several classes of binaries containing luminous stars orbiting quiescent compact objects in AU-scale orbits, including white dwarfs (WDs; \citealt{Shahaf2024MNRAS, Garbutt2024MNRAS}), neutron stars (NSs; \citealt{El-Badry2024OJAp_ns,ElBadry2026}), and black holes (BHs; \citealt{El-Badry2023MNRAS_bh1, El-Badry2023MNRAS_bh2, Chakrabarti2023AJ}). These AU-scale compact object binaries were underrepresented in observed samples prior to Gaia but have turned out to be relatively common \citep[e.g.][]{Yamaguchi2025PASP, Yamaguchi2026arXiv}.

While detailed case studies of individual systems can reveal possible evolutionary pathways (e.g., \citealt[]{El-Badry2023MNRAS_bh1, El-Badry2023MNRAS_bh2, El-Badry2024OJAp_ns1}), population-level studies that model the rates and distribution of observed properties across a sample can test how frequently different pathways occur and how evolutionary outcomes vary with system properties (e.g., \citealt{Hallakoun2024ApJL, Nagarajan2025PASP_2, Yamaguchi2025PASP, Yamaguchi2026arXiv}). This allows us to develop empirical models of binary evolution that can be generalized to make predictions for other populations, beyond what is possible from individual systems alone. However, to conduct such population-level studies, it is necessary to account for both the selection function of the observed samples and uncertainties in their parameters.

\begin{figure*}
    \centering
    \includegraphics[width=0.98\linewidth]{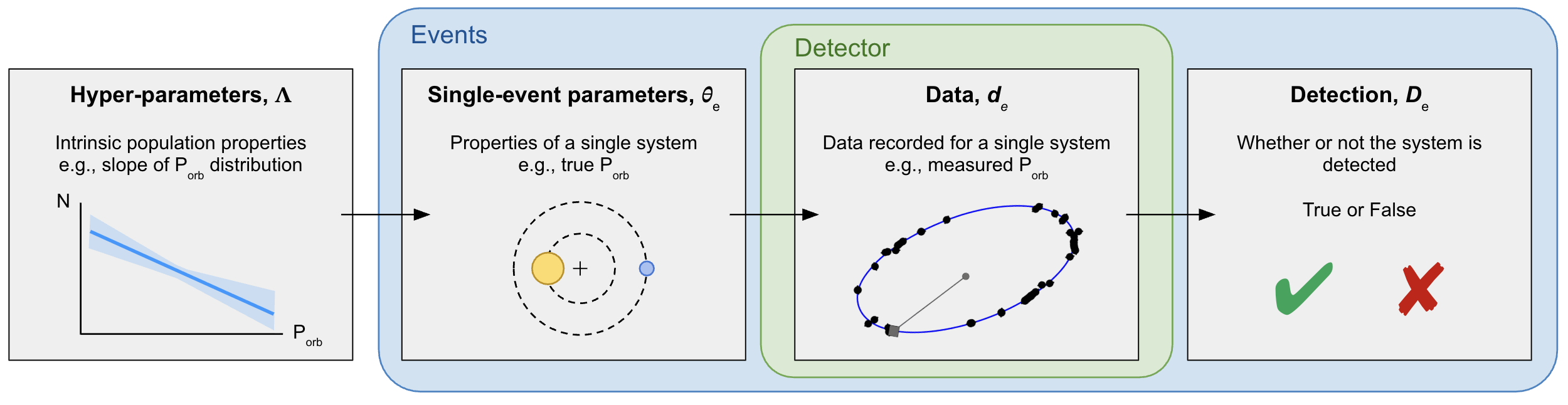}
    \caption{Schematic of our hierarchical Bayesian inference model.}
    \label{fig:population_model}
\end{figure*}

Hierarchical Bayesian modeling is one method for inferring population statistics from incomplete and noisy survey data \citep[e.g., ][]{Loredo2004AIPC, Mandel2009ApJ, Foreman-Mackey2014ApJ, Mandel2019MNRAS}. In particular, it has been employed extensively by the gravitational wave community to model compact binary coalescences (CBC) detected by the LIGO-Virgo-KAGRA collaboration \citep[e.g., ][]{Wysocki2019PhRvD, Fishbach2018ApJL, Abbott2016PhRvX, Abbott2019ApJL, Abbott2021PhRvX, Abbott2023PhRvX, 2026ApJ..1005L..51A}. The goal of this method is to infer population-level parameters (or ``hyperparameters") that define the \textit{intrinsic} distributions from which individual systems are drawn as well as the overall rate. Each of these systems has a set of observables that are constrained by noisy measurements. Only a fraction of all systems are detected at sufficiently high significance, forming a set of detected systems. The term ``hierarchical" refers to this structure of constraints on population parameters via constraints on system parameters from data (Figure \ref{fig:population_model}).

While hierarchical Bayesian modeling is extensively used in interpreting gravitational wave observations, it has yet to be widely applied to model electromagnetic observations of compact objects in the Milky Way (see \citealt{Farr2011ApJ} as an early example). This is in part due to the difficulty of constructing the selection functions of observed electromagnetic samples, which are complex and heterogeneous. The sample of Gaia astrometric orbits, however, is large, homogeneously constructed, and amenable to population modeling. \citet{El-Badry2024OJAp} showed that the selection function for Gaia astrometric orbits can be accurately approximated. By accounting for the Gaia scanning law and implementing a realistic noise model, their model simulates astrometric measurements for a given binary and then fits these measurements through a processing pipeline which closely resembles the Gaia pipeline. \citet{El-Badry2024OJAp} apply this technique to a simulated population of Galactic binaries to construct a mock catalog of astrometric orbital solutions, which they show reproduces the key properties of the true catalog.  

In this work, we apply hierarchical Bayesian modeling to a sample of compact object binaries compiled from the Gaia astrometric orbit catalog.  Most of our analysis is focused on a large sample of WDMS binaries from \citet{Shahaf2024MNRAS} (\citetalias{Shahaf2024MNRAS} henceforth), implementing results from \citet{El-Badry2024OJAp} for the selection function. This sample is well-suited for population inference with flexible models, as it contains over 3000 homogeneously selected, high-confidence systems -- about $100\times$ and $1000\times$ larger than the number of NS and BH binaries in the same catalog, respectively. Given that these binaries primarily host carbon-oxygen (CO) WDs in AU-scale orbits, they are likely products of mass transfer (MT) from AGB donors that avoided a classic common envelope phase and its associated dramatic orbital shrinkage. The details of this interaction are still poorly understood (\citetalias{Shahaf2024MNRAS}; \citealt{Belloni2024A&A, Yamaguchi2024PASP, Yamaguchi2025PASP, Yamaguchi2026arXiv}). Constraints on the intrinsic population demographics of these binaries can thus serve as a valuable benchmark for models of AGB MT. Recently, \citet{Yamaguchi2025PASP} and \citet{Yamaguchi2026arXiv} (\citetalias{Yamaguchi2026arXiv} henceforth) forward-modeled this sample, deriving a simulated WDMS binary population that can reproduce key observed properties after accounting for selection effects. These studies enabled direct constraints on binary mass transfer, but relied on qualitative comparisons with the observed sample and did not formally account for measurement uncertainties. Here, we complement their work by using hierarchical Bayesian modeling to formally infer intrinsic population properties directly from the data, while remaining agnostic to evolutionary models. The derived intrinsic distributions of orbital and stellar parameters, as well as the overall rate, will be valuable as empirical bias-corrected benchmarks for physical models to reproduce. In particular, our approach enables inference of population properties even in regions where few systems are observed and/or detectability is low. Such features can be difficult to probe with qualitative comparisons between observed and simulated samples, but can provide key tests of evolutionary models in regimes where their predictions are most uncertain.

The remainder of this paper is organized as follows. In Section \ref{sec:maths}, we provide a mathematical overview of hierarchical Bayesian inference. In Section \ref{sec:s24_sample}, we summarize the work of \citetalias{Shahaf2024MNRAS} to construct the observed sample of WDMS binaries from the astrometric orbit catalog. In Section \ref{sec:method}, we describe key components of our method, including the steps to model the selection function and parameterization of the underlying population model. In Section \ref{sec:mcmc_fitting}, we detail the MCMC fitting procedure and in Section \ref{sec:results}, we provide the results. We describe the inferred intrinsic properties, compare them to the results of the forward model from \citetalias{Yamaguchi2026arXiv}, and discuss implications of our findings to binary evolution in Section \ref{sec:discussion}. We test our method on the current sample of two Gaia BHs and 27 Gaia NSs in Section \ref{sec:bh_ns}. Finally, we summarize our main findings in Section \ref{sec:conclusion}.

\section{Mathematical background} \label{sec:maths}
  
Here, we review the key equations that underlie our hierarchical Bayesian model. This section closely follows the works of \citet{Essick2024ApJ} and \citet{Essick2025PhRvD}, which readers can consult for detailed derivations. We summarize the variables and their definitions in Table \ref{tab:variables}. The relationship between variables is illustrated in the schematic of Figure \ref{fig:population_model}. 

\begin{table*}[ht]
    \centering
    \caption{Summary of key variables of the framework of hierarchical Bayesian inference introduced in Section \ref{sec:maths}.}
    \begin{tabular}{|c|c|c|}
         \hline
        Variable & Definition & Comments \\
        \hline \hline
        $\theta$ & Single-system parameters & Inferred: $P_{\rm orb}$, $M_1$, $M_{\rm WD}$, eccentricity \\
        &&Fixed: [Fe/H], age, RA, Dec, PMRA, PMDec, $\varpi$, orientation \\  \hline
        $\Lambda$ & Population (hyper-)parameters & e.g. slope of power-law orbital period distribution \\  \hline
        $\mathcal{K}$ & Astrophysical rate & Intrinsic number of systems in the volume considered \\  \hline
        $\Lambda_{\rm inj}$ & Injection parameters &  Assumed distributions of system parameters in the injection set \\  \hline
        $d_e$ & Data recorded for the system & e.g. estimated astrometric parameters \\  \hline
        $D_e$ & Detection flag for the system & 1 if system is detected, 0 otherwise \\  \hline
        $P(D\mid\Lambda)$ & Detection probability for a population & The fraction of systems that enter the observed sample \\  \hline
        $P(D\mid\theta)$ & Detection probability of a single system & Encodes information about the selection function \\  \hline
    \end{tabular}
    \label{tab:variables}
\end{table*}

Each system (i.e., an astrometric binary, or in the case of gravitational wave data, a compact object merger) is assumed to be generated through an inhomogeneous Poisson process, whether or not it was observed. In other words, the population is described by an expected density in parameter space, and counts in disjoint regions of this space are independent. Under this assumption, the intrinsic rate of systems, $dN/d\theta$, is given by $\mathcal{K}\,p(\theta\mid\Lambda)$, where $\theta$ is the set of system parameters (e.g. orbital period, eccentricity, component masses), $\mathcal{K}$ is the total number of astrophysical systems (e.g. the space density of WDMS binaries), $\Lambda$ is the set of population parameters (i.e. parameters describing the mass, period, and eccentricity distributions), and $p(\theta\mid\Lambda)$ is the probability distribution from which each system is drawn. In our analysis, we drop explicit time dependence and assume that the properties of the underlying population remain constant over the observing period considered. Our model of the selection function (Section \ref{ssec:selection_func}) accounts for changes in scanning frequency over the course of the Gaia mission. 

The ``detection probability," which represents the fraction of all systems that are detected, can be written as: 
\begin{align} \label{eqn:det_prob}
    P(D |\Lambda) & = \int d\theta \, p(\theta|\Lambda)P(D|\theta)
\end{align} 
where $D$ is a discrete variable denoting detection, and $P(D|\theta)$ encodes the selection function (i.e., detection probability of a particular system), which should itself be interpreted as an integral over realizations of noisy data, $d$: $P(D|\theta) = \int dd P(D|d) p(d|\theta)$~\citep{Essick2024ApJ}. This integral can be calculated using a Monte Carlo approximation: 

\begin{align} 
    \begin{split}
    P(D |\Lambda) 
    &= \int d\theta \, p(\theta|\Lambda_{\rm inj})\left[\frac{p(\theta|\Lambda)}{p(\theta|\Lambda_{\rm inj})}P(D|\theta)\right] \\
    &\approx \hat{P}(D |\Lambda) \equiv \frac{1}{N_{\rm inj}}\sum_{k}^{N_{\rm inj}}\left[\frac{p(\theta^{(k)}|\Lambda)}{p(\theta^{(k)}|\Lambda_{\rm inj})}D^{(k)}\right] 
    \end{split}
\end{align} 

where $D^{(k)}$ is either 0 or 1 depending on whether or not a system is detected. The bracketed term inside the sum is referred to as the ``importance weight." $N_{\rm inj}$ draws are taken from a reference ``injected" distribution, which can take any functional form and is parameterized by population parameters $\Lambda_{\rm inj}$. 
The choice of injected population can affect the precision of the Monte Carlo estimate, but the integral will eventually converge for any injected distribution with enough samples as long as it has support over all $\theta$ where systems could be detected. The $N_{\rm inj}$ draws are typically referred to as ``injections'' or ``an injection set.''

The Poisson likelihood, $\mathcal{L}$, of detecting $N_{\rm det}$ systems (i.e., WDMS binaries in the observed sample) with data $d_e$ (i.e., astrometric parameters and uncertainties) is $\mathcal{L}\propto \exp(-N_{\rm det}^{\rm exp})\prod_{e=1}^{N_{\rm det}}\left.\frac{dN_{\rm det}}{dd}\right|_{d=d_e}
$, where $N_{\rm det}^{\rm exp}=\mathcal{K}P(D\mid\Lambda)$ is the expected number of detections (Equation \ref{eqn:det_prob}). The differential expected number of detections is given by $\frac{dN_{\rm det}}{dd}\big|_{d=d_e}
=\mathcal{K}\int d\theta_ep(\theta_e\mid\Lambda)p(d_e\mid\theta_e)$ where we marginalize over the true parameters $\theta_e$ of each system. Substituting these expressions into the Poisson likelihood and taking its logarithm yields:
\begin{multline}
    \log(\mathcal{L}) = N_{\rm det}\log(\mathcal{K}) - \mathcal{K} P(D |\Lambda) \\
        + \sum_e^{N_{\rm det}} \log \left[ \int d\theta_ep(\theta_e \mid \Lambda) p(d_{e} \mid \theta_e)  \right]
\end{multline}
Qualitatively, the first term rewards models that predict enough systems to account for the number of detected systems, the second term penalizes models that predict too many detections, and the final term quantifies how well the predicted population is able to explain the single-system parameters of the detected sample.  

In the special case of perfect measurements (i.e., negligible uncertainties), $p(d_{e} \mid \theta_e)$ becomes a delta function:
\begin{align}
    \int \, d\theta_e\,p(\theta_e \mid \Lambda)  \delta(\theta_e - \hat{\theta}_e) = p(\hat{\theta}_e \mid \Lambda) 
\end{align}

Otherwise, the integral can be approximated with a Monte Carlo sum, where $N_\mathrm{mc}$ samples are drawn from 
a reference posterior distribution constructed using the system likelihood and a chosen reference prior, $\theta_e^{(k)} \sim p(\theta|d_e,\Lambda_\mathrm{ref}) \propto p(d_e|\theta) p(\theta|\Lambda_\mathrm{ref})$. The reference prior, parameterized by $\Lambda_{\rm ref}$, can take any functional form. In this case: 
\begin{equation} \label{eqn:measurement_uncert}
    \int \, d\theta_e\, p(\theta_e \mid \Lambda)  p(d_e\mid\theta_e) \propto \frac{1}{N_{\rm mc}} \sum_{k}^{N_{\rm mc}} \frac{p\left(\theta_e^{(k)} \mid \Lambda \right)}{p\left(\theta_e^{(k)} \mid \Lambda_{\rm ref} \right)}
\end{equation}

We draw samples of the system parameters in Gaia’s Thiele–Innes coordinates assuming flat reference priors, and use these samples to calculate $M_{\rm WD}$ through a nonlinear transformation. This transformation introduces a Jacobian term in $p(\theta|\Lambda_\mathrm{ref})$, resulting in a non-unity importance weight. We detail this calculation in Appendix \ref{appendix:ref_prior}. 

\section{Gaia WDMS astrometric binaries} \label{sec:s24_sample}

Based on the astrometric orbital solutions alone, we do not know \textit{a priori} the nature of the component stars. In this section, we briefly summarize the work of \citetalias{Shahaf2024MNRAS} to isolate systems likely hosting MS stars orbiting compact objects. 

\subsection{AMRF} \label{ssec:amrf}

The selection of compact object binaries hinges on a quantity called the ``astrometric mass ratio function" (AMRF), first introduced by \citet{Shahaf2019MNRAS}. Written in terms of observables, 
\begin{align}
    \mathrm{AMRF} = \frac{\alpha}{\varpi} \left(\frac{M_1}{M_{\odot}} \right)^{-1/3} \left(\frac{P_{\rm orb}}{\rm yr} \right)^{-2/3}
\end{align}
where $\alpha$ is the angular photocentric semi-major axis, $\varpi$ is the parallax, $M_1$ is the mass of the photometric primary, and $P_{\rm orb}$ is the orbital period. This is calculated for every astrometric binary in the observed astrometric catalog. For $M_1$, \citetalias{Shahaf2024MNRAS} takes values from the \texttt{gaiadr3.binary\_masses} table \citep{GaiaCollaboration2023A&A_b}. 

The AMRF can also be written in terms of the (photometric secondary-to-primary) mass ratio, $q$, and flux ratio, $\mathcal{S}$: 
\begin{align}
    \mathrm{AMRF} = \frac{q}{(1 +q)^{2/3}} \left( 1-\frac{\mathcal{S}(1+q)}{q(1+\mathcal{S})} \right)
\end{align}
Depending on the nature of the secondary, the relation between $q$ and $\mathcal{S}$ will vary. In the simplest case of a completely dark secondary, $\mathcal{S}=0$ and we have a one-to-one relation between AMRF and $q$. Using stellar models, \citetalias{Shahaf2024MNRAS} derived AMRF$-q$ relations for different possible secondaries, and by considering the highest possible AMRF for a given $M_1$, they obtained boundaries in AMRF$-M_1$ space which define three regions: 
\begin{itemize}
    \item Class-I: secondary consistent with a single MS star.
    \item Class-II: secondary inconsistent with a single MS star, consistent with an inner binary of two MS stars (i.e., a triple).
    \item Class-III: secondary inconsistent with both a single MS star and an inner binary of two MS stars, likely to be a dark compact object (most commonly, a WD). 
\end{itemize}
For each system, \citetalias{Shahaf2024MNRAS} run a Monte Carlo experiment in which they draw $10^5$ samples of the measured parameters from uncorrelated Gaussian distributions.
They calculate the corresponding AMRFs for each sample, from which they derive the probability that the system belongs to each of the three classes. While isolating class-III systems would lead to the purest sample of compact objects, it would exclude most WDs around MS stars above $\sim0.7\,M_{\odot}$, where the primary is too bright for an inner MS binary to be ruled out. They therefore consider both class-II and class-III systems, which together they call the ``non-class I" (NCI) sample, with a class-I probability (Pr(I)) below 10\%. They use a less stringent color-based cut to discard potential triples (Section \ref{ssec:color_excess}). They also set an upper bound on the primary mass of $1.2\,M_{\odot}$ to within $1\sigma$, as more massive stars are unlikely to enter the NCI sample. 

\subsection{Color excess} \label{ssec:color_excess}

While an inner binary of two low-mass MS stars may contribute little to the total light, it can make the system appear significantly redder than expected for a single MS component.

\citetalias{Shahaf2024MNRAS} calculate the ``color excess" (CE) of every system in the NCI sample, defined as the difference between the observed $(B-I)$ color and the expected $(B-I)$ color for the MS primary alone: 
\begin{equation}
    \mathrm{CE} = (B-I)_{\rm observed} - (B-I)_{\rm expected}
\end{equation}
If the system is redder than expected, $\mathrm{CE}>0$, suggesting it may be a triple. 

They use the \texttt{stam} code to obtain $(B-I)_{\rm expected}$ by interpolating PARSEC stellar evolutionary tracks on metallicity and absolute V-band magnitude at a fixed age of $2\,$Gyr. They take observed B, V, and I-band magnitudes from Gaia's \texttt{synthetic\_photometry\_gspc} catalog and use metallicities derived by \citet{Zhang2023MNRAS}. To account for uncertainties, they take $10^4$ Monte-Carlo samples of CE for each system in the NCI sample by drawing the G-band magnitude and metallicity from an uncorrelated bivariate Gaussian distribution to calculate the probability, $\rm Pr(red)$, that $\mathrm{CE}>0$ (i.e., the fraction of Monte-Carlo draws that pass $\mathrm{CE}>0$). They isolate those where $\mathrm{Pr(red)}< \mathrm{Pr(red)_{max}} = 64\%$ to form the final ``no color excess" (NCE) sample.

\subsection{Sample purity} \label{ssec:purity}

A sample that is significantly contaminated by systems not accounted for in the underlying population model will yield biased and unreliable results. Previous work suggests that the observed sample from \citetalias{Shahaf2024MNRAS} is mostly pure, although the CE calculation may introduce small levels of contamination. We summarize their findings here. 

First, \citet{Yamaguchi2024PASP} conducted spectroscopic follow-up for a subset of astrometric binaries in the NCI sample and found that the majority of their Gaia orbital solutions were consistent with the measured radial velocities. 

Meanwhile, \citet{Yamaguchi2025PASP} studied the same sample using a forward-modeling approach. 
Their work showed that the method used by \citetalias{Shahaf2024MNRAS} to isolate WDMS binaries is effective, with the cuts on AMRF and CE removing the vast majority of contaminating MSMS binaries and triples, respectively. In fact, they found that the CE cut may be more stringent than needed and likely removes a significant fraction of WDMS binaries in the NCI sample. 

However, the update to the forward model from \citetalias{Yamaguchi2026arXiv} revealed evidence of a systematic error in the measured [Fe/H] derived by \citet{Zhang2023MNRAS} used to calculate predicted colors for the observed sample. This leads to a negative offset in CEs for a large fraction of systems, which can only be replicated in the simulated sample with an added ``correction" to the [Fe/H] values. In doing so, \citetalias{Yamaguchi2026arXiv} find that the fraction of contaminating triples in the observed NCE sample may be larger than previously estimated. 

In light of this, we test stricter cuts on CE (i.e. smaller $\rm Pr(red)_{max}$; Section \ref{ssec:color_excess}) to construct smaller, but presumably purer, observed samples. We apply the same cuts to our injection set and run our population inference (Section \ref{sec:method}). We provide details in Appendix \ref{appendix:prredmax_test}. In short, a stricter CE cut ultimately yields a lower inferred number density and a steeper decline towards lower primary masses, both of which are consistent with reduced contamination from triples. However, because all best-fit values are within $\sim 2\sigma$ of each other for $\rm Pr(red)_{max} = 20-64\%$, we conclude that our results are largely insensitive to this choice. Unless otherwise stated, we therefore retain $\rm Pr(red)_{max} = 64\%$. 


\section{Methods} \label{sec:method}

\subsection{Injection distributions} \label{ssec:inj_dist}

\begin{figure*}
    \centering
    \includegraphics[width=0.98\linewidth]{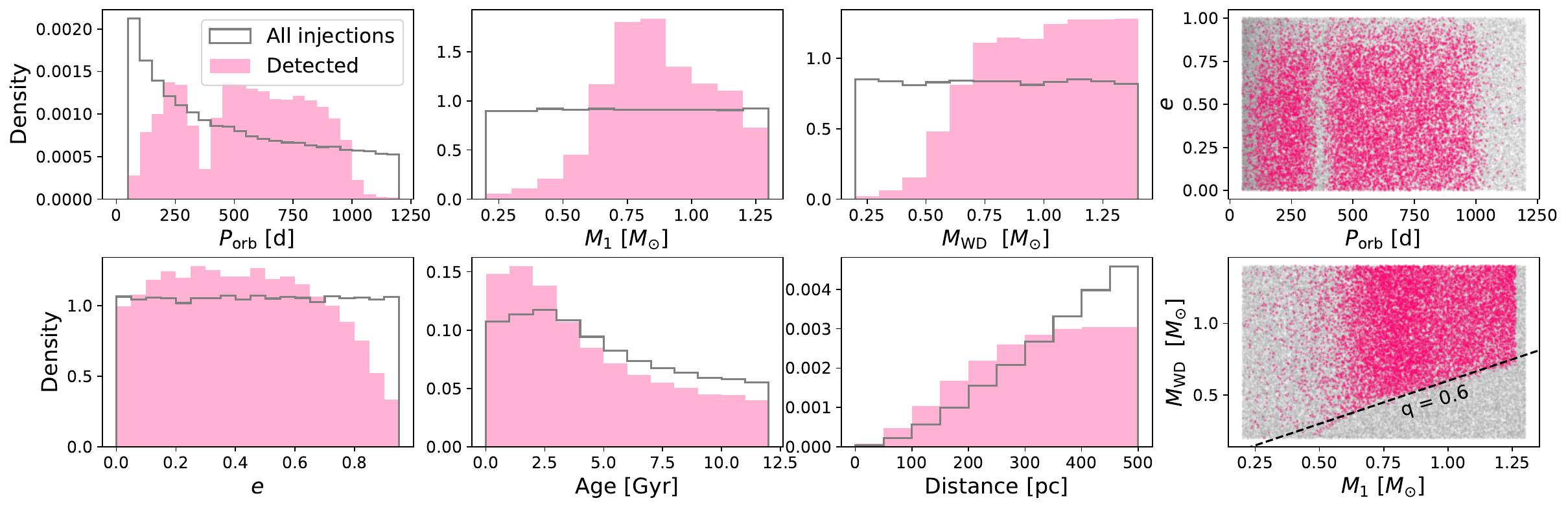}
    \caption{1D histograms (normalized) and 2D scatter plots of parameters of the injection set. Across all panels, we plot the complete injection set in gray and detected systems in pink (Section \ref{ssec:inj_dist}). $P_{\rm orb}$, $M_1$, $M_{\rm WD}$, and $e$ are the parameters fit in our model, while the distributions of age and distance are assumed to be known. The difference between the injected and detected samples are a result of the various steps in the selection pipeline (Section \ref{ssec:selection_func}), as discussed in the text. The black dashed line in the $M_{\rm WD}-M_{1}$ plot marks a mass ratio of 0.6.
    }
    \label{fig:injections}
\end{figure*}

We model distributions of four key population parameters: orbital period ($P_{\rm orb}$), eccentricity ($e$), primary mass ($M_1$), and WD mass ($M_{\rm WD}$). All other parameters, to which the selection function is less sensitive or whose distributions are expected to be better understood, are assumed to have known distributions. 

We summarize the injection distributions below: 
\begin{itemize}
    \item $P_{\rm orb}$: Power law, $\propto P_{\rm orb}^{-0.5}$, between $50-1200\,$d 
    \item $M_1$: Uniform between $0.2 - 1.3\,M_{\odot}$
    \item $M_{\rm WD}$: Uniform between $0.2 - 1.4\,M_{\odot}$
    \item Eccentricity ($e$): Uniform between $0 - 1$
    \item $[\rm Fe/H]$: Sampled from a skew normal distribution fitted to metallicities from \citet{Zhang2023MNRAS} (with \texttt{quality\_flags}$<8$). 
    \item Age: Uniform between the minimum and maximum ages (with an upper bound of 12 Gyr) for the primary to be on the MS given $M_1$ and [Fe/H]. We evaluate the bounding ages by interpolating MIST evolutionary tracks \citep{Morton2015ascl, Choi2016ApJ}. 
    \item Sky position (RA, Dec), proper motion (PMRA, PMDec), parallax ($\varpi$): Sampled from a Milky Way model generated using \texttt{Galaxia} \citep{Sharma2011ApJ}. We set an upper bound on the distance of $500\,$pc and model only systems within this volume.
    \item Remaining orbital elements (time and argument of periapsis, longitude of the ascending node, inclination): Random orientations
\end{itemize}
For a sufficiently large injected population, the final results are insensitive to the assumed distributions. In practice, however, it is preferable that the injected population is not too different from the true population, as this minimizes the number of injections required for precise Monte Carlo estimates. The injected population should also have support in all regions of parameter space where systems could plausibly be detected. 

Apparent G-band magnitudes are predicted for the injected population using MIST synthetic photometry (assuming a completely dark secondary), and extinctions are obtained using the \texttt{combined19} dust map \citep{Drimmel2003A&A, Marshall2006A&A, Green2019ApJ} from the \texttt{mwdust} package \citep{Bovy2016ApJ}. 

\subsection{Modeling the selection function} \label{ssec:selection_func}

We generate $N_{\rm inj} = 10^5$ WDMS binaries for the injection population. Each of these must go through the following series of steps, which model the selection function to construct the observed sample in \citetalias{Shahaf2024MNRAS}: 
\begin{enumerate}
    \item Using \texttt{gaiamock}\footnote{https://github.com/kareemelbadry/gaiamock} \citep{El-Badry2024OJAp}, we determine whether each binary receives an astrometric orbit in Gaia DR3.
    \item For each binary, generate $10^5$ instances of AMRF using the fitted orbital parameters and their uncertainties to calculate $\mathrm{Pr(I)}$. If $\mathrm{Pr(I)}< 10\,\%$, the system enters the NCI sample. 
    \item Use \texttt{stam} to interpolate PARSEC models to calculate CE as the difference between the ``observed" $(B-I)$ color of the MS star given its true parameters and its ``expected" color given a fixed age of $2\,$Gyr and absolute V-band magnitude. We generate $10^4$ Monte Carlo realizations of CE to compute $\rm Pr(red)$. If $\rm Pr(red) < 64\%$ (Section \ref{ssec:purity}), the system enters the final NCE sample, and $D_e = 1$. 
\end{enumerate}

The details of \texttt{gaiamock} can be found in \citet{El-Badry2024OJAp}. In short, for a given binary, the code applies the Gaia scanning law to generate mock epoch astrometry that accounts for orbital motion and Gaia's finite spatial resolution \citep{Lindegren2022}, and implements an empirical noise model \citep{Holl2023A&A}. This is then processed through a pipeline that replicates the one used to construct the observed astrometric catalog \citep{Halbwachs2023A&A}, in which the mock astrometry is fit with increasingly complex orbital models until a sufficiently good fit is obtained, after which final quality cuts are applied to remove potentially poor fits. 

For in-depth descriptions of steps 2 and 3, we refer readers to Section 3 of \citet{Yamaguchi2025PASP}, which largely replicates the analysis of \citetalias{Shahaf2024MNRAS} (Section \ref{ssec:amrf}). However, we follow the updated CE calculation as detailed in Appendix B of \citetalias{Yamaguchi2026arXiv}, where a systematic error in the reported [Fe/H] values from \citet{Zhang2023MNRAS} was identified as largely responsible for the observed trend between CE and $M_1$ (Section \ref{ssec:purity}). Following \citet{Yamaguchi2025PASP}, for $M_1$, we use the true injected value with a fixed uncertainty of $0.06\,M_{\odot}$ (the mean of observed values), rather than reproducing the full calculation used to construct the \texttt{binary\_masses} catalog. Therefore, while we approximately account for scatter in the $M_1$ estimates, but we do not capture potential systematic biases or correlations between those estimates and the fitted orbital parameters.

Figure \ref{fig:injections} shows 1D histograms and scatter plots of several parameters of the injection set. In pink, we highlight the $\sim 13,400$ injections ($\sim13\%$ of the total) that are detected, pass all cuts, and enter the NCE sample.

The features in the $P_{\rm orb}$ distribution are primarily introduced by the detectability of astrometric binaries. There is a dearth of systems with $P_{\rm orb}\sim 1\,$yr due to degeneracy between parallactic and orbital motion. Sensitivity also falls off steeply below $\sim 100\,$d, where astrometric wobbles become too small to be detected, and above $\sim 1000\,$d, where orbits do not get sufficient phase coverage. For the latter reason, \citetalias{Shahaf2024MNRAS} imposes an upper limit of $P_{\rm orb}=1000\,$d, which explains the sharp cutoff. None of the detected systems have measured periods above $1000\,$d, but a few have true periods above this cutoff, which have been underestimated by the Gaia pipeline. Astrometry is also biased against highly eccentric systems, which spend most of their time near apocenter, leading to sparse sampling of their orbits. 

Another notable feature is the lack of detected systems with mass ratios below $\sim 0.6$. This is due to the AMRF cut, which requires a sufficiently massive secondary to rule out a single MS secondary. We also note a fall in detectability for ages below $\sim 2\,$Gyr. This stems from how CE is calculated in \citetalias{Shahaf2024MNRAS}, where the expected color is determined using stellar models at a fixed age of $2\,$Gyr. As solar-type stars are predicted to become bluer on the MS before becoming redder as they approach the sub-giant branch, this biases the sample against the youngest systems. In Appendix \ref{appendix:triang_age}, we change the assumed age distribution to one that peaks at $2\,$Gyr. We find this only results in subtle changes to the final parameter distributions, but because it slightly raises the average detection efficiency, it decreases the inferred space density by a factor of $\sim 1.4$. Since our main conclusions are unaltered, we do not perform a more thorough test of the age distribution in this work. 

Lastly, overestimated CEs partly explain the low detection efficiency at $M_1 < 0.6\,M_{\odot}$, in addition to the fact that fainter sources are less likely to receive orbital solutions. We further describe such effects from the various steps in the processing pipeline in Section \ref{ssec:pcc_in_steps}. 

\subsection{Population models} \label{ssec:pop_model}


We make the simplifying assumption that the joint probability distribution of the four fitted parameters is separable: 
\begin{align}
    \begin{split}
    p(\theta|\Lambda) &= p(P_{\rm orb}, M_1, M_2, e|\Lambda)   \\ &=p(P_{\rm orb}|\Lambda)p(M_1|\Lambda)p(M_2|\Lambda)p(e|\Lambda)
    \end{split}
\end{align}
This is equivalent to assuming that any correlation between parameters in the observed sample stems from the selection function and is not a property of the underlying population.

We parameterize the individual distributions as follows:
\begin{itemize}
    \item $P_{\rm orb}$: single power law with a slope $\alpha_{\rm Porb}$
    \item $M_{\rm 1}$: broken power law with slopes $\alpha_{\rm M1,lo}$ and $\alpha_{\rm M1,up}$ below and above a single break at a location $M_{1,0}$ 
    \item $M_{\rm WD}$: broken power law parameterized by $\alpha_{\rm MWD,lo}$, $\alpha_{\rm MWD,up}$, $M_{\rm WD,0}$
    \item $e$: broken power law parameterized by $\alpha_{\rm e,lo}$, $\alpha_{\rm e,up}$, $e_{\rm 0}$
\end{itemize}
In total, with the addition of the rate $\mathcal{K}$, there are a minimum of 11 population parameters to be fitted. To guide our selection of parameterizations, we used predictions of binary evolution models. \citet{Yamaguchi2025PASP} (and later, \citetalias{Yamaguchi2026arXiv}) start with relatively well-understood parameter distributions for MSMS binaries to generate the zero-age population, which they then evolve using simple analytic models into present-day WDMS binaries. We use their predicted intrinsic distributions (see e.g. Figure 14 of \citealt{Yamaguchi2025PASP}) to motivate the choices described above.

An important caveat is that if the chosen parameterization is unable to fit the true underlying distribution, then the resulting population parameters are of limited value, regardless of the implementation of the method. However, a posterior predictive check (Section \ref{sec:results}) lets us simulate the observed/detected sample from the inferred best-fit population model, which we can then compare with the true data. If the distributions of predicted observables agree well with reality, it suggests the chosen distributions were sensible (though not necessarily unique). 
For recent discussions in the gravitational-wave context, see e.g., \citet{Abbott2021ApJL, Romero-Shaw2022PASA, 2023ApJ...955..107F, Miller:2026buq}.
Additionally, by verifying agreement between predicted and observed 2D distributions, we can check that the separability assumption does not induce a noticeable systematic bias. We leave correlated models for future work.

\subsubsection{Residual B-spline fitting} \label{sssec:bspline}


While Appendix \ref{appendix:parametric} shows that the simple parametric models alone successfully reproduce the overall trends of the observed NCE sample, they fail to reproduce observations in the tails of the distributions. This is unsurprising, as the fit is largely driven by where most observed systems lie, so an inflexible model is most sensitive to the shape of the distribution near the peak and will tend to extrapolate to the tails. 
Some massive WDs ($\gtrsim 0.8\,M_{\odot}$) may be merger products or neutron stars. The inferred space density at the upper tail of the WD mass distribution may thus provide valuable information to compare with independent constraints from other WD samples. 


Motivated by this, we incorporate an additional component to account for deviations from the single-break power-law model for the $M_{\rm WD}$ model. We use a basis spline (B-spline) function for this component, which provides flexibility in shape and degree of complexity. For the following discussion, we closely follow Appendix C.1 in \citet{2026ApJ..1005L..51A}, to which we refer readers for more details (see also \citealt{Edelman:2021zkw,2023ApJ...946...16E, Godfrey:2023oxb}). A B-spline function has the following form: 
\begin{align}
    S(x) = \sum_{i=0}^{N_b-1} c_jB_{i, k;t}(x)
\end{align}
where $N_b$ is the number of basis vectors considered, $B_{i, k;t}$ are the basis functions of order $k$ (and degree $n=k-1$) and knots $t$, and $c_j$ are the spline coefficients. 

This becomes an additional term in the population model: 
\begin{align}
    \log p(\theta\mid\Lambda)= \log p_{\rm base}(\theta\mid\Lambda) + S(M_{\rm WD}) - \log \rm C
\end{align}
where $p_{\rm base}$ is the single-break power law model as before, and $C$ is the normalization constant that ensures that $p(\theta\mid\Lambda)$ integrates to 1.

By default, we use cubic splines ($k = 4$) and set $N_b=10$. We have tested $N_b =5-15$ and find that our results are qualitatively insensitive to this choice. The number of knots is given by $N_b-k-1$, and we evenly space their locations linearly in $M_\mathrm{WD}$, bounded by $0.2$ and $1.4\,M_{\odot}$. The fitted parameters are the spline coefficients: $c_j$. 

To combat overfitting, we add a smoothing prior that penalizes large differences between neighboring spline coefficients:
\begin{align}
    p(\textbf{c}\mid\tau) = \exp\left(-\frac{1}{2} \tau \textbf{c}^T \mathcal{D}^{T}_r\mathcal{D}_r\textbf{c}\right)
\end{align}
where $\textbf{c}$ is the vector of coefficients, $\mathcal{D}_r$ is the r-order difference matrix (we use $r = 2$), and $\tau$ is a parameter that controls the level of smoothing. Following \citet{2026ApJ..1005L..51A}, we set $\tau = 5$. We have tested values of $\tau = 1-10$ and found no significant differences in our results, in particular on the inferred rate of massive WDs (Section \ref{ssec:binary_evol}). We also impose ridge penalties on the coefficients for added stability. 

\subsection{Incorporating measurement uncertainties} \label{ssec:mc_samples}

As described in Section \ref{sec:maths}, hierarchical Bayesian inference allows us to incorporate measurement uncertainties in the parameters of the observed systems into the fit (Equation \ref{eqn:measurement_uncert}). 

For each WDMS binary, we draw samples of the Thiele-Innes orbital elements from a multivariate Gaussian using the covariance matrix derived from the correlation vector in the \texttt{gaia\_source} table, which are then converted to Campbell elements. These Gaussian draws approximate the measurement likelihood in Gaia's native Thiele–Innes coordinates. We transform the resulting samples into physical orbital parameters via non-linear transformations, accounting for the corresponding Jacobian in the importance weights, resulting in non-Gaussian likelihoods in the orbital parameters (Appendix \ref{appendix:ref_prior}). Samples of $M_1$ are drawn from a split normal distribution, taking the lower and upper uncertainties quoted in the \texttt{binary\_masses} table. Lastly, $M_{\rm WD}$ is calculated using $P_{\rm orb}$, $M_1$, parallax, and photocentric semi-major axis. 

For the BHs and NSs (Section \ref{sec:bh_ns}), we take reported uncertainties from the literature and draw samples assuming independent Gaussian distributions. 

\section{MCMC fitting} \label{sec:mcmc_fitting}

\begin{figure*}
    \centering
    \includegraphics[width=0.9\linewidth]{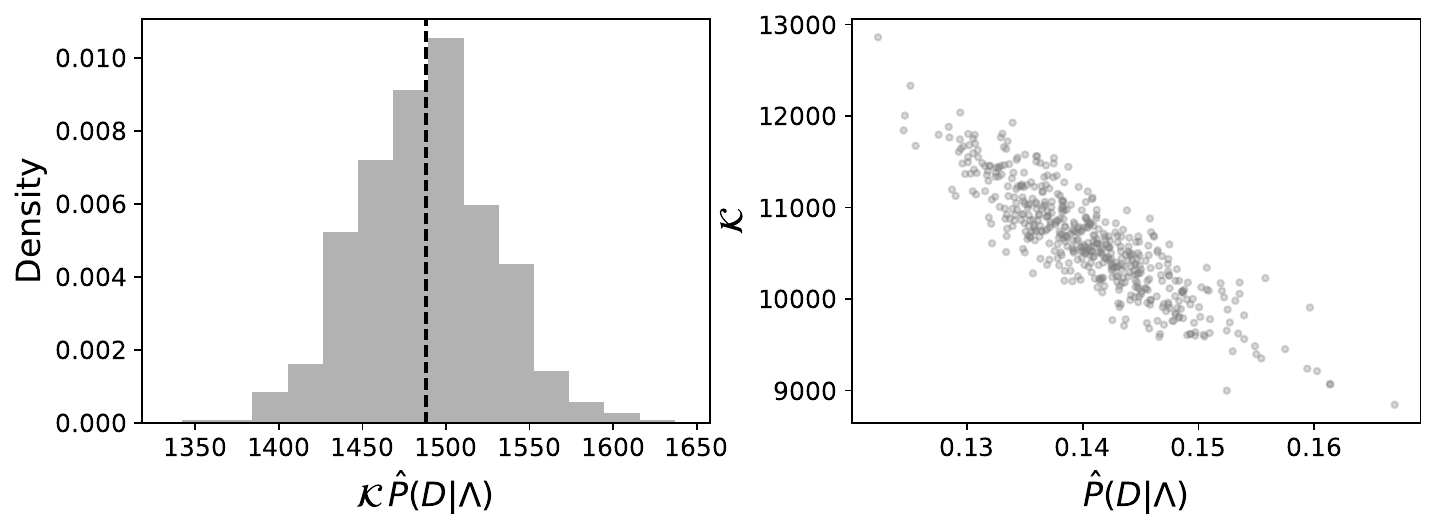}
    \caption{\textit{Left}: Distribution of $\mathcal{K}\hat{P}(D\mid\Lambda)$, the expected number of detected systems, for 500 random draws of $\Lambda$ from the posterior. 
    Their peaks line up with the number of observed systems in the \citepalias{Shahaf2024MNRAS} catalog within 500 pc (vertical dashed line). \textit{Right}: $\hat{P}(D\mid\Lambda)$ vs. $\mathcal{K}$. As expected, they are anti-correlated, making their product roughly constant.}
    \label{fig:rate}
\end{figure*}

Finally, we run an MCMC fit with the package \texttt{emcee} \citep{Foreman-Mackey2014ApJ}. We isolate observed systems from \citetalias{Shahaf2024MNRAS} with parameters within the range of the injection set (Section \ref{ssec:inj_dist}), yielding a total of 1488 systems to which we fit our models. 

We impose flat priors on all fitted hyper-parameters. Based on the results of \citetalias{Yamaguchi2026arXiv}, we ensure that all single-break power laws are peaked (i.e., $\alpha_{\rm lo} > 0$ and $\alpha_{\rm hi} < 0$). Otherwise, a wide range of values is allowed for each parameter. As described in Section \ref{sssec:bspline}, we incorporate a smoothing prior for the B-spline coefficients of the $M_{\rm WD}$ component.\footnote{The code will be made publicly available on GitHub upon publication.}

In Appendix \ref{appendix:corner}, we show the resulting corner plot for our fiducial model as a visual diagnostic of good convergence. After an initial 1000-step  burn-in and 2000 additional equilibration steps, and we run 3000 production steps with 400 walkers for a total of $1.2\times10^6$ samples. We have also checked that the trace plots show no obvious trends or clustering, suggesting well-mixed chains. Moreover, we have tested our code on a mock intrinsic population and have confirmed that it recovers the input distributions. 

\section{Results} \label{sec:results}

We summarize the best-fit population parameters in Table \ref{tab:pop_params}. We provide the full MCMC samples of all parameters, including the B-spline coefficients, as supplemental material in the online version of this manuscript. 

In Figure \ref{fig:rate}, we plot the distribution $\mathcal{K}\hat{P}(D\mid\Lambda)$ for 500 draws of the population parameters, $\Lambda$, and the rate, $\mathcal{K}$, from the posterior. As expected, the peak is near the total number of WDMS binaries in the observed sample. In the right panel, we plot $\mathcal{K}$ against $\hat{P}(D\mid\Lambda)$ for the same draws. There is an anti-correlation, as required for their product to remain near the observed value.

\begin{table}[]
    \centering
    \caption{Inferred parameters for our fiducial model. The quoted values are medians with 16th--84th percentile intervals from the one-dimensional marginal posteriors.}
    \begin{tabular}{|c| c| c|}
        \hline
         Parameter & Prior & Median [16, 84]\% \\
         \hline \hline
         $\alpha_{\rm Porb}$ & $U(-2, -0.0001)$ & -0.445 [-0.4918, -0.3962] \\
        \hline
        $M_{1, 0}$ & $U(0.2, 1.3)$ & 1.07 [1.05, 1.08] \\
        $\alpha_{\rm M1,lo}$ & $U(0, 50)$ & 0.413 [0.256, 0.573] \\
        $\alpha_{\rm M1,hi}$ & $U(-50, 0)$ & -33.8 [-44.8, -21.6] \\
        \hline
        $M_{\rm WD, 0}$ & $U(0.2, 1.4)$ & 0.583 [0.581, 0.585] \\
        $\alpha_{\rm MWD,lo}$ & $U(0, 50)$ & 10.1 [6.7, 13.2] \\
        $\alpha_{\rm MWD,hi}$ & $U(-50, 0)$ & -21.8 [-24.8, -18.8] \\
        \hline
        $e_{0}$ & $U(0, 1)$ & 0.0459 [0.0453, 0.0467] \\
        $\alpha_{\rm e,lo}$ & $U(-1, 15)$ & 8.36 [6.81, 10.11] \\
        $\alpha_{\rm e,hi}$ & $U(-50, 0)$ & -2.14 [-2.19, -2.08] \\
        \hline
        $\log(\mathcal{K})$ & $U(-50, 50)$ & 9.27 [9.22, 9.32] \\
        \hline
    \end{tabular}
    \label{tab:pop_params}
\end{table}

\begin{figure*}
    \centering
    \includegraphics[width=0.98\linewidth]{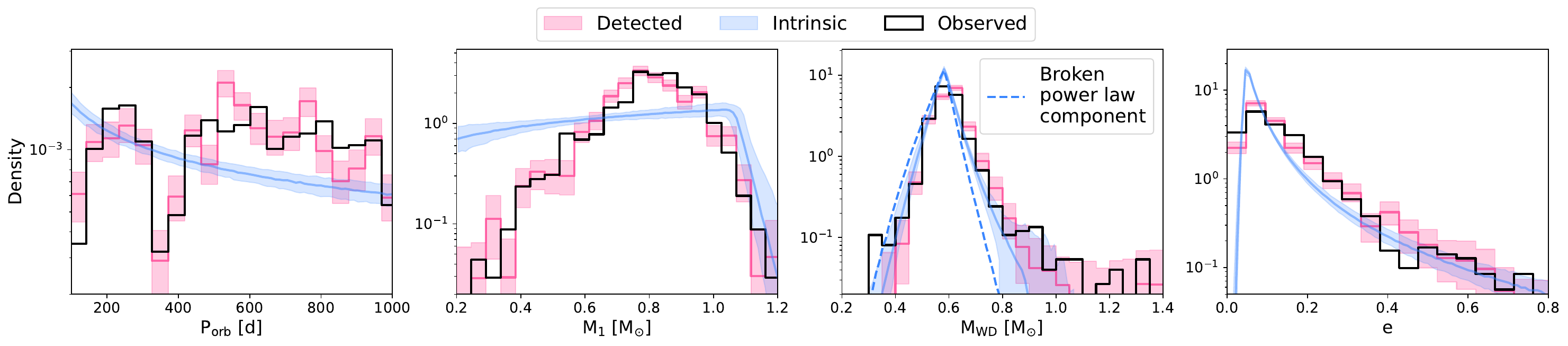}
    \includegraphics[width=0.98\linewidth]{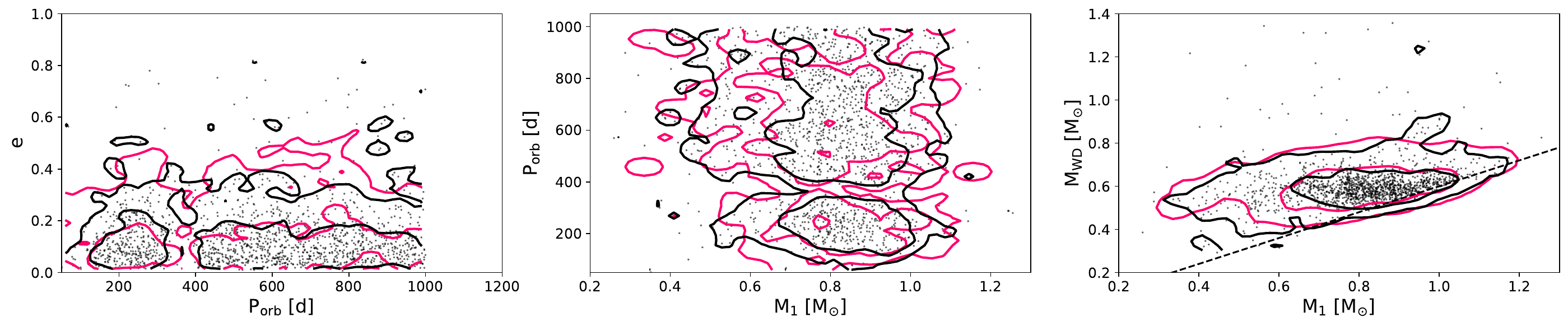}
    \caption{
    \textit{Top}: 1D distributions of the fitted parameters.
    The observed sample from \citetalias{Shahaf2024MNRAS} is plotted in black and compared with the PPD in pink (i.e., the distribution of maximum-likelihood estimates as returned by \texttt{gaiamock} for our injection campaign reweighed to follow the inferred astrophysical distribution).
    The inferred distribution of the intrinsic population is shown in blue; note that this is a distribution over a different variable (true parameters) than the pink and black distributions (maximum-likelihood estimates of parameters). }
    The shaded regions cover the 5th-95th percentiles.
    As expected, the PPD (pink) and observed (black) distributions largely overlap.
    \textit{Bottom}: 2D histograms, where two contours correspond to the 68th and 95th percentiles of the cumulative probability.
    We also plot the individual data points for the observed sample.
    \label{fig:ppc}
\end{figure*}


\begin{figure*}
    \centering
    \includegraphics[width=0.99\linewidth]{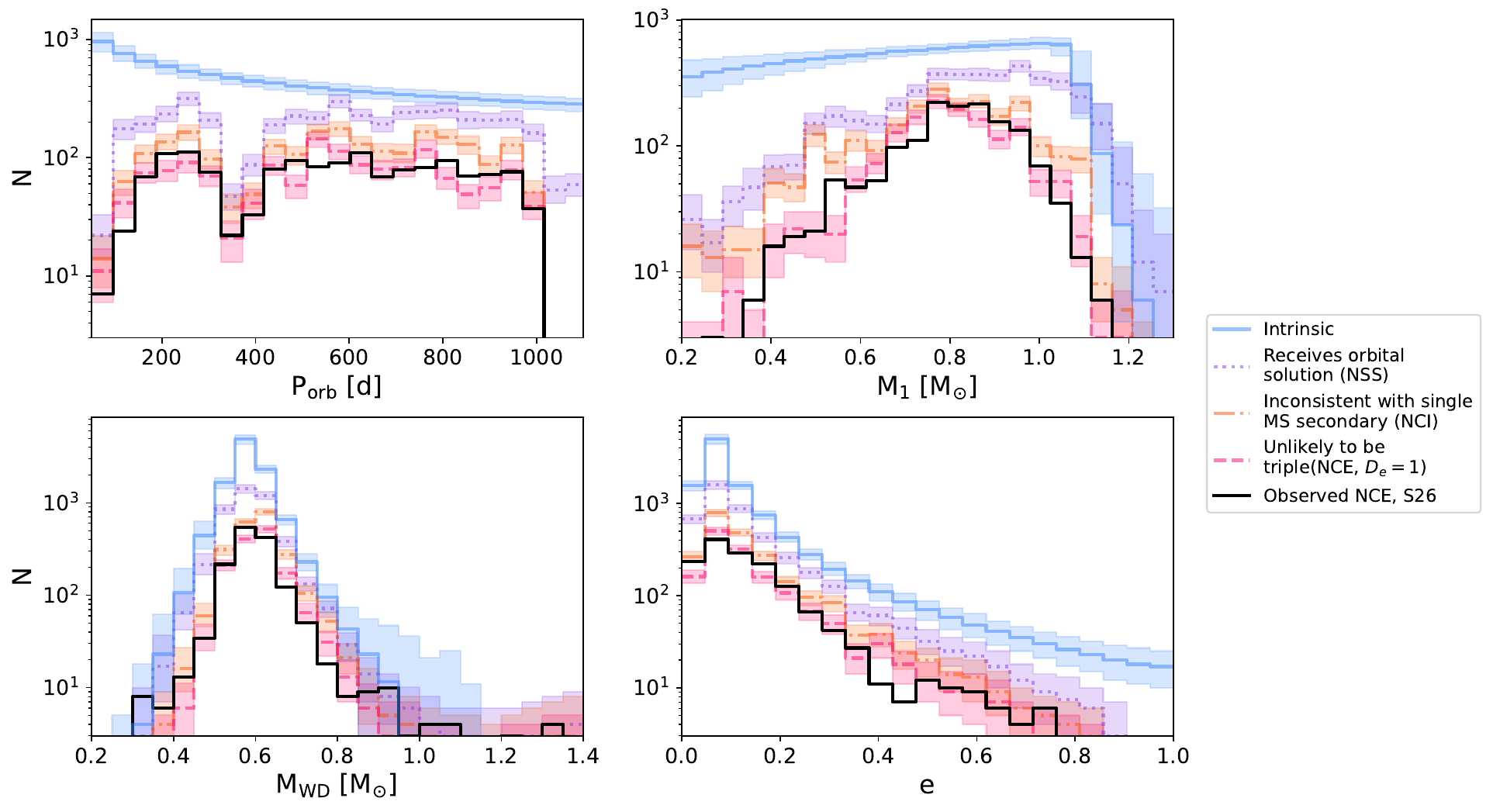}
    \caption{Distribution of the fitted parameters after each step in the processing pipeline (Section \ref{ssec:selection_func}), starting off with $N =\mathcal{K}$ systems drawn from the best-fit population model.}
    \label{fig:ppc_2}
\end{figure*}

To verify that our results are reasonable, we conduct a posterior predictive check (PPC) by comparing the predicted and observed parameter distributions.  

For this, we take 500 random draws of $\Lambda$ and $\mathcal{K}$ from the posterior. For each draw, we sample the number of detections as $N_{\rm det}\sim\mathrm{Poisson}[\mathcal{K}\hat{P}(D\mid\Lambda)]$ and sample that many systems from the detected injections with probabilities proportional to their importance weights. This yields 500 instances of would-be-detected systems (the posterior predictive distribution; PPD), which we compare with the true NCE sample from \citetalias{Shahaf2024MNRAS} in Figure \ref{fig:ppc}. For these detected systems, we take their ``measured" orbital parameters from \texttt{gaiamock} and draw $M_1$ from a Gaussian around the true value with a fixed uncertainty (Section \ref{ssec:selection_func}), which we then use to calculate the dark companion mass, or $M_{\rm WD}$. The top row shows 1D histograms of the fitted parameters. For comparison, we also overplot the underlying population model where, as expected, we see the assumed power-law forms (with the exception of $M_{\rm WD}$ that has an added B-spline component; Section \ref{sssec:bspline}). In general, the true data and our model agree well. While some bins show discrepancies at the level of the plotted uncertainties, these discrepancies depend in part on the binning and reflect the limited flexibility of our chosen parametric population models. These plots therefore demonstrate that our model successfully reproduces broad population trends but is not intended to capture small-scale substructure in the distributions. The bottom row shows 2D histograms with contours corresponding to the 68th and 95th percentiles. The contours largely overlap with the model, which reproduces the bulk trends in the data. This suggests that the separability assumption in our population model (Section \ref{ssec:pop_model}) does not induce a noticeable systematic bias. In other words, we find that observed correlations primarily arise from the selection function and do not suggest strong correlations between the fitted parameters in the intrinsic population. 


\section{Discussion} \label{sec:discussion}

\subsection{From intrinsic to observed: effects of each selection criterion} \label{ssec:pcc_in_steps}

In Figure \ref{fig:ppc_2}, we apply the same approach as in the PPC (Figue \ref{fig:ppc}) to obtain parameter distributions at every step of the selection pipeline (Section \ref{ssec:selection_func}). The y-axis is the absolute number of systems within $500\,$pc. As expected from Figure \ref{fig:rate}, the inferred occurrence rate of WDMS binaries leads to a final detected sample (pink dashed) that is similar in size to the true dataset (black solid). 

Going from the intrinsic population (blue solid) to the astrometric catalog (purple dash-dot), more massive MS stars are favored because they are brighter. Narrowing down further to the non-class I sample (orange dotted), the AMRF cut preferentially removes more massive MS stars and less massive WDs, for which it is more difficult to rule out luminous secondaries. Lastly, we find that the CE cut preferentially removes low-mass MS stars (orange dotted to pink dashed). This reflects shortcomings of the CE cut: while intended to remove triples, it also removes true WDMS binaries, particularly those with $M_1 \lesssim 0.6\,M_{\odot}$. Indeed, there are a few hundred observed systems, the majority with $M_1 \lesssim 0.7\,M_{\odot}$, whose orbits alone are inconsistent with a triple (i.e., Class-III; Section \ref{ssec:amrf}) yet fail the CE cut from \citetalias{Shahaf2024MNRAS}.

We calculate the space density as the total number inferred, $\mathcal{K}$, divided by the effective volume enclosed within the distance considered, $d_{\rm max} = 500\,$pc. The effective volume is defined as (see Equation 10.9 of \citealt{El-Badry2021PhDT}): 
\begin{equation}
    \tilde{V}\left(d_{\rm max}\right) = 2\pi\int_0^{d_{\rm max}} e^{-z/h_z} \left( d_{\rm max}^2 - z^2 \right) dz
\end{equation}
where we take $h_z = 300\,$pc to be the scale height of the disk. Taking 1000 instances of $\log(\mathcal{K})$ from the posterior, we calculate a midplane space density of $34932 \pm 1902\,\mathrm{kpc}^{-3}$ of WDMS binaries (within the parameter space probed by the observed sample and the injection set; Section \ref{ssec:inj_dist}). 

\subsection{Comparison to \citetalias{Yamaguchi2026arXiv}} \label{ssec:y26_comparison}

\begin{figure*}
    \centering
    \includegraphics[width=0.99\linewidth]{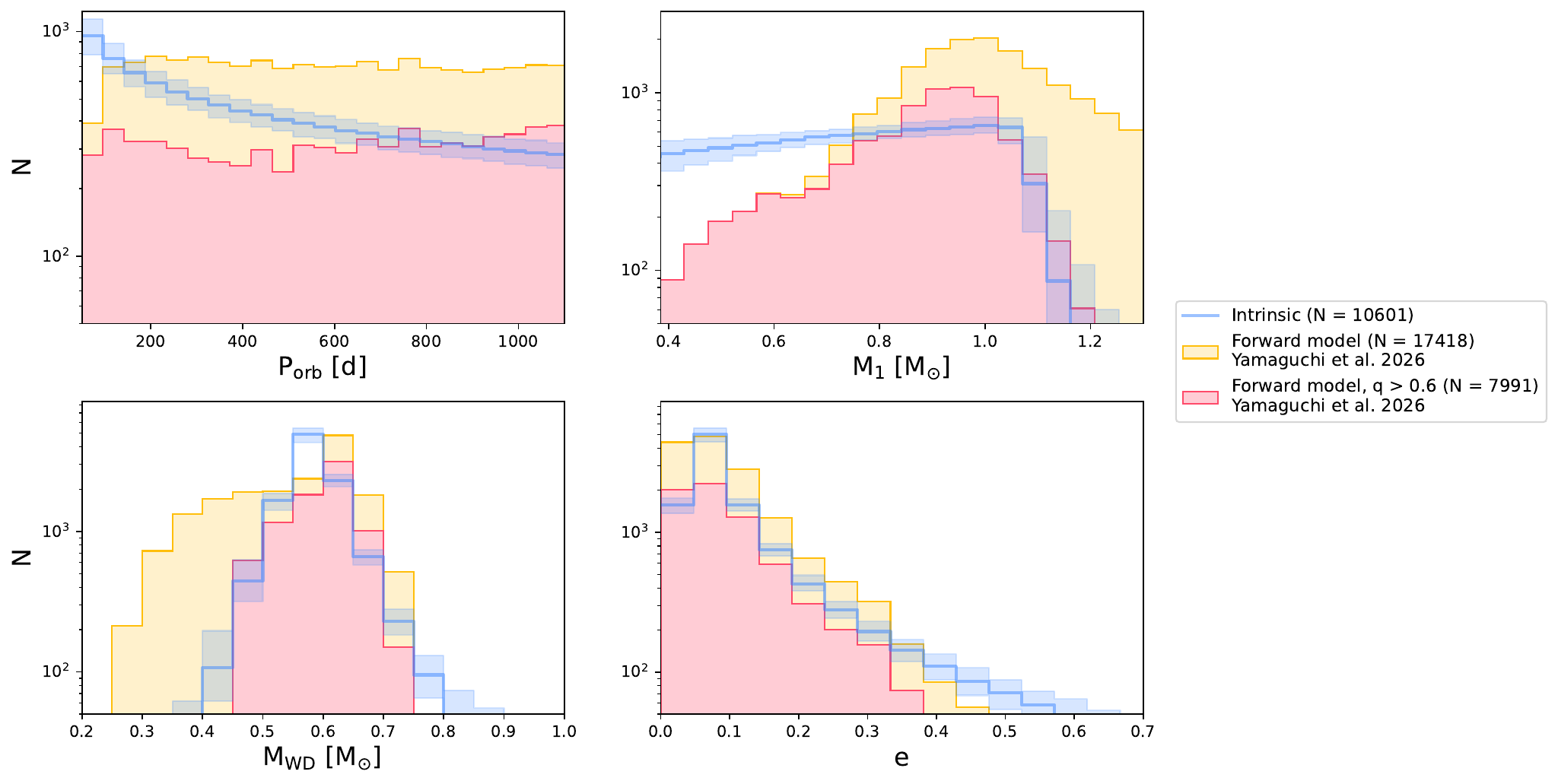}
    \caption{Comparison of the intrinsic parameter distributions of WDMS binaries derived in this work and those of the simulated population in \citetalias{Yamaguchi2026arXiv}. In pink, we isolate systems from \citetalias{Yamaguchi2026arXiv} with $q > 0.6$, to which the observed astrometric sample is confined. There is rough agreement between our results and the $q > 0.6$ subset, in the total numbers as well as the shapes of the $M_{\rm WD}$ and $e$ distributions. In contrast, our model predicts a significantly larger fraction of systems with $P_{\rm orb} \lesssim 600\,$d and a shallower decline in $M_1$ below the break.}
    \label{fig:ppc_3}
\end{figure*}

In Figure \ref{fig:ppc_3}, we compare our results to those from \citetalias{Yamaguchi2026arXiv}, who closely follow the forward modeling approach taken by \citet{Yamaguchi2025PASP} with a few updates to the assumed binary evolution and selection function, but with qualitatively similar results for $P_{\rm orb}=100-1000\,d$. In yellow, we plot all WDMS binaries in their simulated population within our injection domain. In pink, we isolate systems with $q = M_{\rm WD}/M_1 > q_{\rm min} = 0.6$, to which the observed sample is sensitive due to the AMRF cut (Figure \ref{fig:injections}).

Overall, our results more closely match those of the $q > 0.6$ subset than the full population from \citetalias{Yamaguchi2026arXiv}. In particular, we find very close agreement with the $q > 0.6$ subset in the steepness of the slope above the peak of the $M_1$ distribution. This slope is highly sensitive to the exact value of $q_{\rm min}$, quickly becoming shallower as it decreases, such that agreement with our result worsens significantly if, e.g., $q_{\rm min} = 0.55$. This reflects the fact that our hierarchical Bayesian inference is insensitive to regions of the parameter space where no systems should be detectable, and none have been detected. 

Similarly, our model does not probe $M_1$ above the upper limit of $1.2\,M_{\odot}$ imposed to construct the observed sample. For a single-break power-law fit, the uncertainty in the slope is determined primarily by the densely sampled region near the peak and therefore does not reflect the lack of data at the highest masses. If, however, we include an additional break at $1.2\,M_{\odot}$, the marginal posterior distribution of the slope above it matches the prior exactly, confirming we gain no information at these masses.  

Meanwhile, our model predicts a steep decline for the $M_{\rm WD}$ distribution below the peak, once again similar to the $q > 0.6$ population from \citetalias{Yamaguchi2026arXiv}. While low-mass WDs paired with sufficiently low-mass MS stars should be detectable (Figure \ref{fig:injections}), there are very few such observed systems, which leads to the model inferring low intrinsic rates. This means our model does not reproduce much of the post-RGB stable MT products. 

Hereafter, we compare our results only to the $q > 0.6$ population from \citetalias{Yamaguchi2026arXiv}. We find close agreement between the two results near the peaks of the $M_{\rm WD}$ and eccentricity distributions. \citetalias{Yamaguchi2026arXiv} predicts fewer massive WDs $\gtrsim 0.7\,M_{\odot}$. This can be explained by their assumption of a single critical mass ratio for stable MT from AGB donors, which leads most donors that form massive WDs to end up in short-period post-common envelope systems. Similarly, \citetalias{Yamaguchi2026arXiv} employed a fixed empirical relation to determine final eccentricities that falls off to the fourth power of initial eccentricity, which was chosen to reproduce the bulk of the observed sample. In reality, the critical mass ratio and final eccentricity likely depend on the initial stellar and orbital properties. The forward model aimed to reproduce the bulk properties of the population while minimizing free parameters, which explains the divergence at the tails of the population. 

Meanwhile, there are more significant deviations for the $P_{\rm orb}$ and $M_1$ distributions. Firstly, we find a steeper slope in $P_{\rm orb}$ compared to the almost flat distribution of \citetalias{Yamaguchi2026arXiv}, though they converge above $\sim 600\,$d. The $P_{\rm orb}$ distribution of \citetalias{Yamaguchi2026arXiv} is largely determined by their assumption that all cases of stable MT are fully non-conservative. The discrepancy may therefore point to more complex orbital evolution during mass transfer.

For $M_1$, we find a much shallower decline below the peak compared to \citetalias{Yamaguchi2026arXiv}. Once again, the relatively sharp drop-off found by \citetalias{Yamaguchi2026arXiv} is due to the constant critical mass ratio imposed for AGB donors, which leads to few low-mass accretors (i.e., present-day MS stars) experiencing stable MT and remaining in sufficiently wide orbits to be detected via astrometry. Naively, our findings are at odds with such a model. To test whether this is a limitation of our chosen single-break power-law model, in Appendix \ref{appendix:m1_models}, we present results of fitting the $M_1$ distribution with two additional models: a double-break power-law model and a single-break power-law model with a residual B-spline component. While these models allow the slope to evolve below the peak, we find that both prefer a relatively flat slope down to $M_1 \sim 0.3\,M_{\odot}$. We thus conclude that hierarchical Bayesian inference suggests that there are more WDs with low-mass companions than predicted by \citetalias{Yamaguchi2026arXiv} and that this is not simply a result of our chosen model. This is further supported by the results of the our fit to the $200\,$pc sample, where there are significantly more observed low-$M_1$ systems (Section \ref{ssec:200pc_vs_500pc}). We speculate that these systems may be post-common-envelope binaries (Section \ref{ssec:binary_evol}) that are difficult to detect because they are faint and the CE cut is biased against them.


Lastly, we find that the total number of systems with $q > 0.6$ predicted by the two works agrees to within $\sim 30\%$. 

We conclude that the qualitative approach taken by \citet{Yamaguchi2025PASP} is advantageous because it incorporates a binary evolution model within the framework, yielding direct constraints on model parameters. It also offers predictions outside the detectability region of the observed sample, where hierarchical Bayesian inference is uninformative. However, their predictive power may fall off at the extremes of the distributions, where few data points exist. Combining the two approaches -- using hierarchical Bayesian inference to fit the underlying present-day population, which is then reproduced from a zero-age binary population by fitting parameters of binary evolution models -- may be a promising avenue to explore in the future. For previous efforts along these lines in related contexts, see \citet{Andrews2018ApJS, Wong:2023}.

\subsubsection{Caveats} \label{sssec:caveats}

While we have shown that our method successfully derives intrinsic population properties consistent with the observed sample, we emphasize that our results are fundamentally limited by the simple parametric models we fit in this work. Most importantly, these models can underestimate uncertainties where few systems are observed. In particular, as we have mentioned in Section \ref{ssec:y26_comparison}, our model confidently predicts very few AU-scale systems with low-mass He WDs and/or massive MS stars, to which the AMRF selection is biased against but which we know exist from other searches that identify hot WDs via photometric UV excess (e.g., \citealt{Yamaguchi2026arXiv}). The uncertainties on slopes are primarily determined by where observed data points are most concentrated and are underestimated away from this. 

On the other hand, parametric models are often more efficient and easier to interpret than more flexible, non-parametric models. This is why we chose them in this work, which served as a first demonstration of the robustness and potential of hierarchical Bayesian inference for modeling samples constructed from the Gaia catalog of astrometric binaries. With this foundation, and with larger sample sizes from future data releases, there is the prospect of increasing model complexity to better characterize regions of parameter space with less sensitivity.

\subsection{500 pc vs. 200 pc sample} \label{ssec:200pc_vs_500pc}

\begin{figure*}
    \centering
    \includegraphics[width=0.98\linewidth]{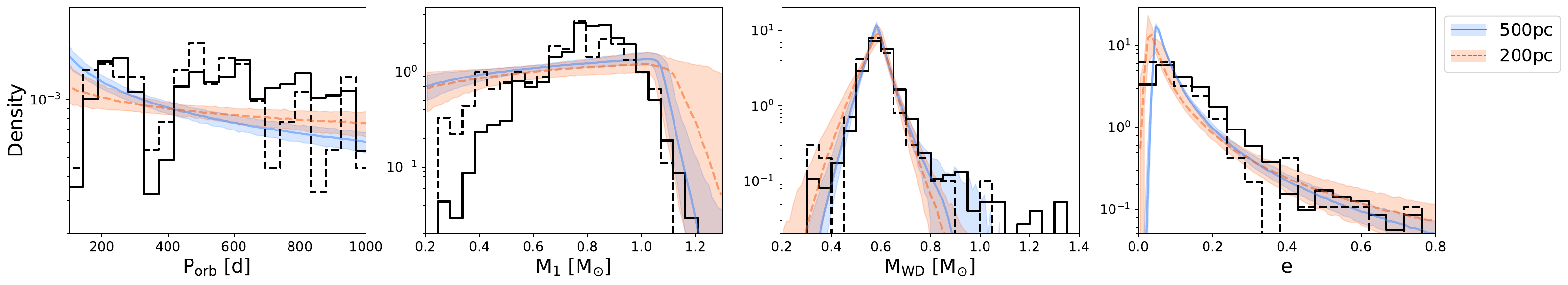}
    \caption{Comparison between the intrinsic parameter distributions obtained from the fit to systems within $500\,$pc (blue) and $200\,$pc (orange). The solid and dashed lines are the distributions of the observed systems within the same distance cuts. The inferred distributions from the two runs are consistent within the plotted 5th to 95th percentile region. In particular, note that the observed $200\,$pc sample contains many more MS stars with $M_1 \lesssim 0.6\,M_{\odot}$, yet the inferred slope below the peak remains flat.}
    \label{fig:200pc_500pc}
\end{figure*}

For our fiducial model, we considered systems within $500\,$pc. To ensure that our results are robust to this choice, we repeat the fitting but with a distance cut of $200\,$pc. In Figure \ref{fig:200pc_500pc}, we compare the inferred intrinsic distributions of the fitted parameters from the two runs. For reference, we also plot the observed distributions. As expected, the significantly fewer observed systems within $200\,$pc (198, compared to 1488 within $500\,$pc) lead to larger uncertainties, particularly at the tails. The best-fit values of all fitted parameters are within $2\sigma$ of each other. The inferred mid-plane space density of $31323 \pm 3723\,\mathrm{kpc}^{-3}$ within $200\,$pc is also consistent with that within $500\,$pc to $< 1\sigma$. 

Additionally, the observed $200\,$pc sample contains significantly more systems with $M_1\lesssim 0.6\,M_{\odot}$ compared to the $500\,$pc sample. In fact, the slope below the $M_1$ peak for the $200\,$pc sample is as shallow as that of the inferred \textit{intrinsic} population from \citetalias{Yamaguchi2026arXiv}, despite the fact that we are biased against low-mass MS stars. This supports our finding of relatively more binaries at low $M_1$ (Section \ref{ssec:y26_comparison}), which is not an obvious result given the full observed sample and highlights a strength of hierarchical Bayesian inference.

\subsection{Implications to binary evolution} \label{ssec:binary_evol}

We briefly discuss our findings in the broader context of binary evolution. 

Firstly, the inferred space density of the AU-scale WDMS binaries studied in this work ($3.5\times10^4\,\mathrm{kpc}^{-3}$) is close to half that of post-common envelope WDMS binaries in much closer orbits ($7.2\times10^4\,\mathrm{kpc}^{-3}$ for $P_{\rm orb}\sim 0.1-2\,$d; \citealt{Shariat2026arXiv}). This highlights that post-AGB MT products that remain in relatively wide orbits are common, and that the lack of these systems discovered pre-Gaia can be attributed to observational biases. 

We find an eccentricity distribution that peaks at a value inconsistent with zero to high significance ($0.046\pm0.001$; Table \ref{tab:pop_params}). As similarly found and discussed in \citet{Yamaguchi2025PASP}, this suggests that circularization of orbits prior to and/or during MT is not as efficient as often assumed (e.g. \citealt{Parkosidis2026A&A_a, Parkosidis2026A&A_b, Parkosidis2026arXiv_c}), or that eccentricity pumping mechanisms, such as those involving an outer tertiary or circumbinary disk (see e.g. \citealt{Naoz2016ARA&A, Lai2023ARA&A} respectively for reviews), play an important role. 

As briefly discussed in Section \ref{ssec:y26_comparison}, we find that a large fraction of systems host low-mass MS stars. Given the sharp peak in the WD mass distribution at $\sim 0.6\,M_{\odot}$ with progenitor masses $\sim 1\,M_{\odot}$, a large fraction of these likely had sufficiently unequal initial mass ratios for mass transfer to become unstable, making them post-common envelope binaries. Indeed, it has been shown that wide orbits can be reproduced for highly evolved AGB donors with very loosely bound envelopes, and/or if extra energy sources, such as recombination energy, aid in the envelope ejection \citep[e.g.,][]{Belloni2024A&A, Yamaguchi2024PASP}. Such systems would not have been predicted by \citetalias{Yamaguchi2026arXiv} who assumed a fixed value for common envelope ejection efficiency calibrated to post-common envelope binaries at much shorter orbital periods.

Moreover, the sharp decline in the WD mass distribution implies intrinsically few He WDs with RGB progenitors around low-mass primaries (Section \ref{ssec:y26_comparison}). If the orbital periods probed correspond to stable mass transfer products from RGB donors, this may be consistent with previous work suggesting that the stability criterion is stricter for RGB than AGB donors (e.g., \citealt{Ge2020ApJ, Temmink2023AA}\citetalias{Yamaguchi2026arXiv}). In other words, the critical mass ratio must be less extreme, meaning more massive accretors (i.e., the present-day MS stars) are required to remain in wide $\sim100-1000\,$d orbits. 

Lastly, we calculate that just $0.97^{+0.55}_{-0.37}\%$ of WDs in the AU-scale binary population have masses $>0.8\,M_{\odot}$. This is significantly smaller than the fraction for samples of field WDs ($\sim 10-20\%$; e.g. \citealt{Tremblay2016MNRAS, Kilic2020ApJ, Jimenez-Esteban2023MNRAS, OBrien2024MNRAS}), as first pointed out by \citet{Hallakoun2024ApJL} and now confirmed by this work, which formally accounts for selection effects. \citet{Hallakoun2024ApJL} suggests that this may be due to the lack of merger products, as a progenitor inner binary may not have been able to evolve given the size of the present-day orbit. Additionally, a $\gtrsim 4\,M_{\odot}$ single star progenitor is needed to form a $\gtrsim0.8\,M_{\odot}$ WD. MT from a more massive donor is more likely to become unstable, leading to a common envelope phase and significant orbital shrinkage; the resulting short-period binaries would not receive astrometric orbits. 

\section{A preliminary test on Gaia BH and NS populations} \label{sec:bh_ns}

We conduct a preliminary test of the code to model astrometric binaries hosting BHs and NSs discovered in Gaia DR3. 

\subsection{Gaia BHs} \label{ssec:bhs}

\begin{figure*}
    \centering
    \includegraphics[width=0.98\linewidth]{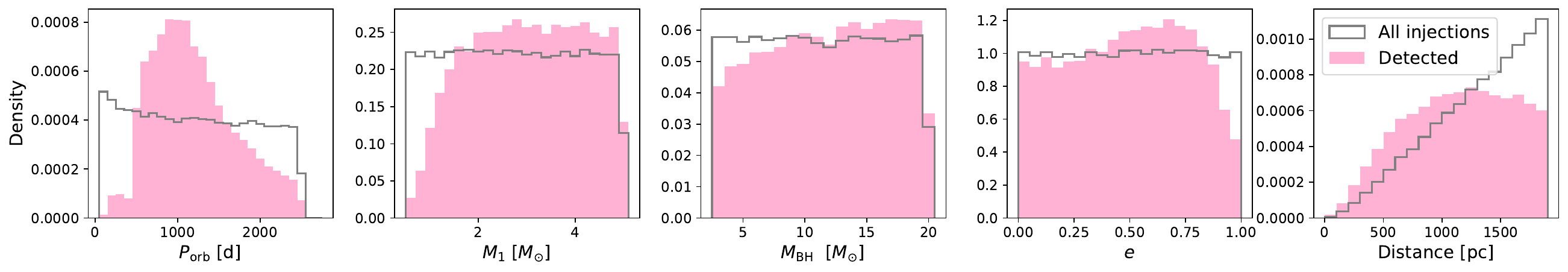}
    \caption{Parameter distributions in the injection set for BHs. In pink, we isolate the ``detected" systems (i.e. those which receive astrometric orbits in DR3).}
    \label{fig:injections_bhs}
\end{figure*}

\begin{table}[]
    \centering
    \caption{Inferred parameters for the BH population, resulting from the fitting of the two BHs in Gaia DR3. Note that the priors on the standard deviations are log-normal.}
    \begin{tabular}{|c|c|c|}
        \hline
         Parameter & Prior & Median [16, 84]\% \\
         \hline \hline
        $\mu(\rm P_{orb})$ & $\mathcal{N}(-0.5, 1.0)$ & -1 [-2, -0.5] \\
        \hline
        $\mu(M_{1})$ & $\mathcal{N}(1.0, 0.2^2)$ & 0.9 [0.8, 1.0] \\
        $\sigma(M_{1})$ & $\mathcal{N}(\ln(0.15), 0.5^2)$ & 0.1 [0.1, 0.2] \\
        \hline
        $\mu(M_{\rm BH})$ & $\mathcal{N}(9.5, 2.0^2)$ & 10 [9, 10] \\
        $\sigma(M_{\rm BH})$ & $\mathcal{N}(\ln(1.0), 0.6^2)$ & 0.8 [0.5, 1.3] \\
        \hline
        $\mu(\rm e)$ & $\mathcal{N}(0.5, 0.15^2)$ & 0.5 [0.4, 0.6] \\
        $\sigma(\rm e)$ & $\mathcal{N}(\ln(0.1), 0.5^2)$ & 0.08 [0.05, 0.13] \\
        \hline
        $\log(\mathcal{K})$ & $U(-50, 50)$ & 3 [2, 5] \\
        \hline
        \hline
    \end{tabular}
    \label{tab:pop_params_bhs}
\end{table}


\begin{figure*}
    \centering
    \includegraphics[width=0.98\linewidth]{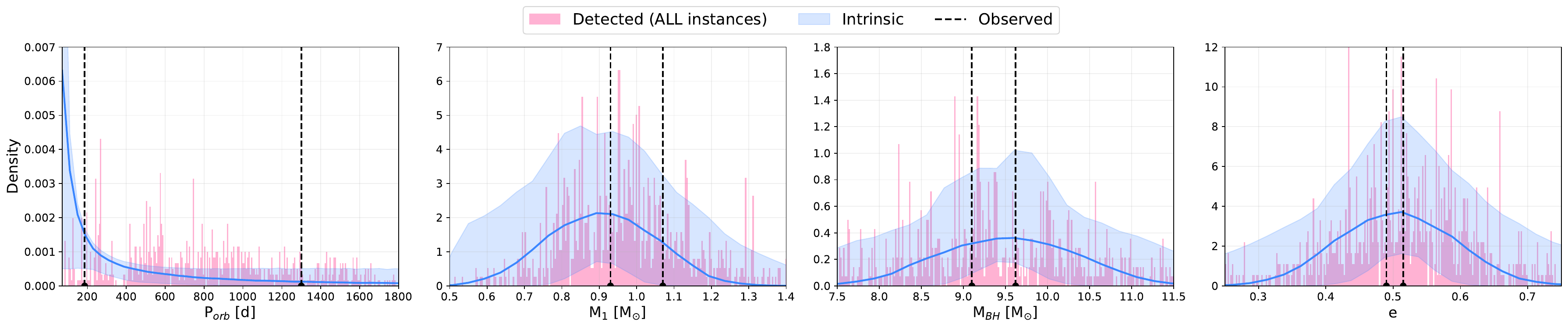}
    \caption{Results of the PPC for the fit to Gaia BHs, analogous to plots in the upper row of Figure \ref{fig:ppc}. However, because each instance in the detected sample has only $\sim 1-3$ systems, we plot the histogram of all systems combined. We mark the locations of the two observed Gaia BHs with dashed vertical lines. 
    }
    \label{fig:1d_ppc_bhs}
\end{figure*}

After extensive searches in the Gaia DR3 catalog \citep{Simon2026ApJ, El-Badry2026arXiv}, only two BHs have been discovered, Gaia BH1 and BH2 \citep{Chakrabarti2023AJ, El-Badry2023MNRAS_bh1, El-Badry2023MNRAS_bh2, El-Badry2026arXiv}, which we assume are all of the BHs present in the catalog. 

For the injection set, we make most of the same choices for distributions as described in Section \ref{ssec:inj_dist}, but extend the lower and upper boundaries to accommodate longer orbital periods ($50 < P_{\rm orb}/\mathrm{d} < 2500$; $\propto P_{\rm orb}^{-0.1}$) and more massive components ($0.5 < M_{1}/M_{\odot} < 5$, $2.5 < M_{\rm BH}/M_{\odot} < 20$). Because we applied no distaPnce cut in the search for BHs, we consider distances up to $2\,$kpc, within which $> 98\%$ of systems in the true astrometric catalog reside. 

We also allow for evolved primaries by setting the upper boundary on age to correspond to the start of the thermally pulsating AGB phase, rather than the end of the main sequence. To avoid overfitting the two data points, we do not implement any B-spline component, keeping only the simple parametric models described in Section \ref{ssec:pop_model}.  

In Figure \ref{fig:injections_bhs}, we plot histograms of the key parameters for systems in the injection set. Since we assume all existing BHs in the dataset have been discovered, the detection here is solely determined by whether a system receives an astrometric orbital solution according to \texttt{gaiamock} (Step 1 in Section \ref{ssec:selection_func}). This is unlike the WDs, for which AMRF is used to identify only a subset that host sufficiently massive WD secondaries relative to their MS primaries. Therefore, the detection for the BHs varies less across the range of component masses considered compared to the WDs (Figure \ref{fig:injections}). There is a bump in detectability at $e \sim 0.4-0.75$, which is absent for the WD injection set. This is a result of the longer orbital periods considered for the BHs, for which sufficiently eccentric orbits with close periastron passage is required to receive orbits \citep{Lam2025_ApJ}. 

With only two data points, we encountered challenges in achieving convergence using the default population models (Section \ref{ssec:pop_model}). To combat this, we replace the broken power laws used to fit $M_1$, $M_{\rm BH}$, and $e$ distributions with truncated Gaussians. This reduces the total number of fitted parameters by three and lets us place more intuitive, informative priors, improving stability and results. We impose normal priors on the means and log-normal priors on the standard deviations. These are summarized in Table \ref{tab:pop_params_bhs}, along with the best-fit values. We verified prior robustness with sensitivity tests in which we varied prior widths and centers over plausible ranges. Still, we caution the readers about over-interpreting the results of this section, in particular with regards to the derived shapes of the distributions which rely on the adopted priors. 

In Figure \ref{fig:1d_ppc_bhs}, we plot results of the PPC. Because each instance has few systems, we plot the combined histogram of all systems across all instances. This allows us to better visualize where most detected systems lie relative to the two Gaia BHs. We confirm that the predicted number of detected systems (i.e. $\mathcal{K}\hat{P}(D\mid\Lambda)$) is $2^{+2}_{-1}$, consistent with the 2 observed. 

In general, uncertainties are much larger than for the WDMS binaries. The intrinsic distributions of component masses and eccentricity peak near the two observed BHs, and detected systems follow similar distributions. This is unsurprising because the selection is relatively insensitive to these parameters over a wide range (Figure \ref{fig:injections_bhs}). As expected, there are many detected systems at the location of each of the observed BHs. 

We predict a total number of $30^{+70}_{-20}$ Gaia BHs-like binaries within $2\,$kpc. While there is a large range, this is consistent with the results of \citet{Lam2025_ApJ} who constructed a model of the selection function of the Gaia astrometric catalog and, using this, calculated $\sim 34 - 57$ BHs orbiting MS stars or red giants with $P_{\rm orb} \sim 100-1500\,$d within $2\,$kpc. Moreover, using binary population synthesis models, \citet{Nagarajan2025PASP_2} estimated that about 1 in 10 million stars in the Milky Way is in an AU-scale orbit around a BH. Estimating the total number of stars within $2\,$kpc as $7.2\times10^8$ (e.g. \citealt{Lam2025_ApJ}), this corresponds to $\sim 72$ BHs. This is also within our estimated range.

With data from the upcoming fourth data release of the Gaia mission, there is a prospect to discover more quiescent BH binaries \citep{GaiaCollaboration2024A&A} and improved population modeling using the framework from this work. 

\subsection{Gaia NSs} \label{ssec:nss}

As a final application, we apply the same hierarchical Bayesian model to the sample of $27$ neutron stars around solar-type stars discovered in {\it Gaia} DR3 \citep[Gaia NSs;][]{El-Badry2024OJAp_ns1, El-Badry2024OJAp_ns, ElBadry2026}.  For simplicity, we once again retain the same parameterization as our fiducial model and exclude the B-spline component. As for the BHs (Section \ref{ssec:bhs}), we emphasize that this is a proof of concept. We will present a dedicated analysis of the Gaia NS population, including alternative parameterizations, natal-kick constraints, and DR4 forecasts, in future work \citep{Shariat2026_GaiaNS}.

\begin{figure*}
    \centering
    \includegraphics[width=0.98\linewidth]{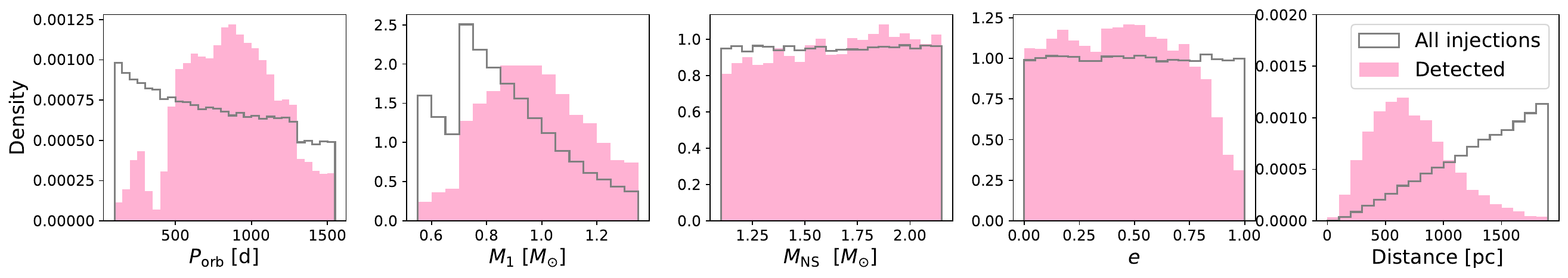}
    \caption{Parameter distributions in the injection set for NSs, analogous to Figure \ref{fig:injections_bhs}.}
    \label{fig:injections_nss}
\end{figure*}

The injection set was adjusted to cover the full observed NS-candidate sample, and contains $140{,}000$ total injections.  
The first $100{,}000$ injections cover $100<P_{\rm orb}/{\rm d}<1300$ and $0.7<M_1/M_\odot<1.4$. 
We add $20{,}000$ injections at $1300<P_{\rm orb}/{\rm d}<1600$ and $0.55<M_1/M_\odot<1.4$, and another $20{,}000$ at $100<P_{\rm orb}/{\rm d}<1300$ and $0.55<M_1/M_\odot<0.7$, to cover the recently discovered Gaia NS candidates \citep{ElBadry2026}. All three sets span $1.1<M_{\rm NS}/M_\odot<2.2$ and $0<e<1$. In Figure \ref{fig:injections_nss}, we plot 1D histograms of several key parameters of the NS injection set.

The compact companion is treated as dark, so the astrometric signal is the photocenter orbit of the visible star. Injections are passed through the {\tt gaiamock} DR3 astrometric-orbit selection. We then apply additional NS-candidate cuts used in the published survey \citep{El-Badry2024OJAp_ns}: $G<15$, $P_{\rm orb}<1600\,{\rm d}$, $M_1<1.35\,M_\odot$, a dark-companion astrometric mass above $1.25\,M_\odot$, a well-measured Thiele-Innes solution, and an astrometric mass-ratio-function cut above the single-main-sequence boundary \citep{Shahaf2024MNRAS}. These cuts capture the bright, high-significance NS candidates sample used for RV validation.

\begin{figure*}
    \centering
    \includegraphics[width=0.98\linewidth]{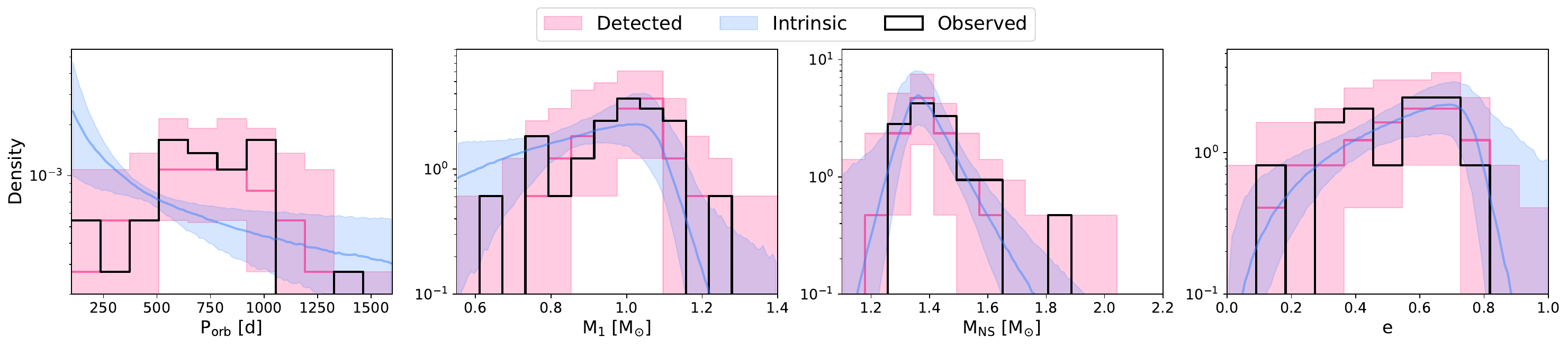}
    \caption{Results of the PPC for the fit to Gaia NSs, analogous to plots in the upper row of Figure \ref{fig:ppc}.}
    \label{fig:1d_ppc_nss}
\end{figure*}

The resulting PPC is shown in Figure~\ref{fig:1d_ppc_nss}. The model reproduces the one-dimensional observed distributions in $P_{\rm orb}$, $M_1$, $M_{\rm NS}$, and $e$ at the level expected for a 27-object sample (see Table \ref{tab:pop_params_nss} for the inferred best-fit parameters).  
The inferred period slope is negative, $\alpha_P=-0.73^{+0.35}_{-0.34}$, implying that the intrinsic population is weighted more strongly toward shorter periods than the detected sample. The visible-star and NS-mass distributions peak near $M_1\approx1.1\,M_\odot$ and $M_{\rm NS}\approx1.4\,M_\odot$, respectively. 
\citet{Schiebelbein-Zwack:2026} fit the original $N=21$ Gaia NS candidates with a two-Gaussian model without correcting for selection effects, finding a narrow component near $1.3\,M_\odot$ and a broader high-mass component. Using the expanded sample of $N=27$ candidates and accounting for selection effects, \citet{Shariat2026_GaiaNS} also find a low-mass peak and a high-mass tail, but infer a broader distribution weighted toward lower masses (see their Section 4).

For the broken-power-law model adopted here, the inferred normalization is $\mathcal{K}=610^{+200}_{-150}$ systems within the modeled 2 kpc volume. The effective selection fraction is $\beta_{\rm eff}=0.043^{+0.010}_{-0.008}$, giving an expected selected count of $\mu=27^{+6}_{-5}$ systems, consistent with the observed $N=27$ sample.

\begin{table}
    \centering
    \caption{Best-fit parameters for the fit to the Gaia NS population.}
    \begin{tabular}{|l|c|c|}
        \hline
        Parameter & Prior & Median [16, 84]\% \\
        \hline \hline
        $\alpha_{\rm Porb}$ & $U(-4, 4)$ & -0.73 [-1.07, -0.38] \\
        \hline
        $M_{1, 0}$ & $U(0.7, 1.4) $ & 1.1 [1.0, 1.1] \\
        $\alpha_{\rm M1,lo}$ & $U(-50, 50)$ & 1.8 [0.5, 4.2] \\
        $\alpha_{\rm M1,hi}$ & $U(-50, 50)$ & -26 [-41, -13] \\
        \hline
        $M_{\rm NS, 0}$ & $U(1.1,2.2)$ & 1.4 [1.3, 1.4] \\
        $\alpha_{\rm MNS,lo}$ & $U(-50, 50)$ & 26 [11, 42] \\
        $\alpha_{\rm MNS,hi}$ & $U(-50, 50)$ & -17 [-22, -13] \\
        \hline
        $e_{0}$ & $U(0, 1)$ & 0.74 [0.69, 0.80] \\
        $\alpha_{\rm e,lo}$ & $U(-50, 50)$ & 1.1 [0.7, 1.6] \\
        $\alpha_{\rm e,hi}$ & $U(-50, 50)$ & -22 [-40, -8] \\
        \hline
        $\log(\mathcal{K})$ & $U(-50, 50)$ & 6.4 [6.1, 6.7] \\
        \hline
    \end{tabular}
    \label{tab:pop_params_nss}
\end{table}


\section{Conclusion} \label{sec:conclusion}

We construct a hierarchical Bayesian inference model to derive intrinsic properties of the WDMS binary population from an observed sample of Gaia DR3 astrometric binaries \citep{Shahaf2024MNRAS}. These systems host WDs with masses $\gtrsim 0.6\,M_{\odot}$ in relatively wide AU-scale separations, making them possible products of stable MT from AGB donors, a channel whose details remain uncertain. We also tested the approach on the smaller samples of NS and BH binaries identified from Gaia DR3. We summarize our main conclusions below: 
\begin{itemize}
    \item \textit{Summary of model performance}: The posterior predictive check demonstrates successful performance, showing that the best-fit intrinsic population forms an inferred detected sample that recovers features of the true observed sample (Figure \ref{fig:ppc}, Section \ref{sec:results}). 
    \item \textit{Population demographics of WDMS binaries}: By accounting for the selection function, our model reveals an intrinsic population with a much wider $M_1$ distribution and higher rates at low $P_{\rm orb}$ compared to the observed sample (Figure \ref{fig:ppc_2}). To summarize, we find a negative power-law slope for $P_{\rm orb}$ ($\alpha_{\rm Porb} \sim -0.4$), a relatively flat $M_1$ distribution between $\sim 0.3-1.0\,M_{\odot}$, $M_{\rm WD}$ distribution that peaks sharply at $\sim 0.6\,M_{\odot}$, and a non-zero peak in eccentricity at $\sim 0.05$. We infer a space density of $\sim 3\times10^4\,\mathrm{kpc}^{-3}$ (Section \ref{ssec:pcc_in_steps}). 
    \item \textit{High rates at short $P_{\rm orb}$ and low $M_1$}: Compared to the predictions of \citetalias{Yamaguchi2026arXiv}, we find a steeper $P_{\rm orb}$ slope and significantly more low-mass primaries with $M_1 \lesssim 0.8\,M_{\odot}$ (Figure \ref{fig:ppc_3}). These discrepancies may reflect oversimplified assumptions about binary evolution in 
    \citetalias{Yamaguchi2026arXiv}, such as fully non-conservative mass transfer, a constant critical mass ratio, and a fixed common envelope ejection efficiency. In particular, the high rate of systems down to $M_1\sim0.3\,M_{\odot}$ combined with the sharp peak at $M_{\rm WD}\sim0.6\, M_{\odot}$ implies relatively extreme initial mass ratios, which may suggest that at least a subset of the population are wide post-common envelope binaries. (Sections \ref{ssec:y26_comparison} and \ref{ssec:binary_evol})
    \item \textit{Limitations of our results}: We note that hierarchical Bayesian inference is insensitive to regions of the parameter space where no systems are expected or detected. Thus, our results are limited to describing the $q > 0.6$ population (Section \ref{ssec:y26_comparison}). Moreover, we caution that our inferred parameter distributions are bounded to our chosen parametric forms with limited flexibility, which may lead to underestimated uncertainties at the tails (Section \ref{sssec:caveats}).
    \item \textit{Massive WDs and eccentric orbits}: The $M_{\rm WD}$ distribution implies just $\sim 1\%$ of massive WDs above $0.8\,M_{\odot}$, significantly lower than the fraction for field WDs. Moreover, the non-zero peak in eccentricity suggests inefficient orbit circularization or eccentricity pumping mechanisms (Section \ref{ssec:binary_evol}). 
    \item \textit{Gaia BH and NS statistics}: We apply our method to fit the sample of the two BHs and 27 NS candidates discovered in Gaia DR3, and derive that $\sim 30$ and $\sim600$ such systems exist within $2\,$kpc (Section \ref{sec:bh_ns}). The relatively small datasets mean significantly larger uncertainties in the derived parameter distributions than those for the WDs. We expect results to improve largely with more discoveries expected in Gaia DR4.
\end{itemize}

\section{Acknowledgments}

We thank the anonymous referee for their constructive feedback which improved this manuscript. We thank Maya Fishbach and Casey Lam for helpful conversations during the early development of this project.  

This research was supported by NSF grants AST-2307232 and AST-2540180.
NY acknowledges support from the Ezoe Memorial Recruit Foundation scholarship.

This research was supported by the Natural Sciences and Engineering Research Council of Canada (NSERC).
R.E. is supported by NSERC Grant RGPIN-2023-03346, and A.M.F. is supported by NSERC Grant DIS-2022-568580.

This research was also supported in part by grant NSF PHY-2309135 to the Kavli Institute for Theoretical Physics (KITP) and made use of the Astrophysics Data System, which is funded by NASA under Cooperative Agreement 80NSSC21M00561.

This work has made use of data from the European Space Agency (ESA) mission {\it Gaia} (\url{https://www.cosmos.esa.int/gaia}), processed by the {\it Gaia} Data Processing and Analysis Consortium (DPAC,
\url{https://www.cosmos.esa.int/web/gaia/dpac/consortium}).
Funding for the DPAC has been provided by national institutions, in particular the institutions participating in the {\it Gaia} Multilateral Agreement.

C.S. acknowledges support from the Department of Energy Computational Science Graduate Fellowship. This material is based upon work supported by the U.S. Department of Energy, Office of Science, Office of Advanced Scientific Computing Research, under Award Number DE-SC0026073. This work made use of \texttt{OverCite} \citep{Shariat2026}, an in-editor citation tool for \LaTeX.

\vspace{5mm}
\facilities{Gaia}

\software{\texttt{isochrones}, \texttt{OverCite}}

\appendix

\section{Importance Sampling with Uncertainties} \label{appendix:ref_prior}

Here, we derive importance weights for Monte Carlo integration to account for system uncertainties (Section \ref{sec:maths}). These weights incorporate the nonlinear transformation from Gaia’s original coordinates to $M_{\rm WD}$ as well as the differences between the reference sampling distributions and the distributions from which the injection set was drawn. 

For each astrometric binary, Gaia provides the Thiele–Innes orbital elements, $P_{\rm orb}$, and $e$, along with their covariance matrix, which we use to generate samples. The primary mass, $M_1$, is provided separately in the \texttt{gaiadr3.binary\_masses} catalog and is sampled from an asymmetric Gaussian defined by its reported lower and upper uncertainties. 

To calculate $M_{\rm WD}$, we must first convert Thiele-Innes to Campbell elements: 
\begin{align}
    A = a_0(\cos\omega \cos\Omega -\sin\omega\sin\Omega\cos i) \\
    B = a_0(\cos\omega \sin\Omega + \sin\omega\cos\Omega\cos i) \\
    F = -a_0(\sin\omega \cos\Omega +\cos\omega\sin\Omega\cos i) \\
    G = -a_0(\sin\omega \sin\Omega -\cos\omega\cos\Omega\cos i) \\
\end{align}
where $a_0$ is the angular photocentric semi-major axis, $\omega$ is the argument of periastron, $\Omega$ is the ascending node, and $i$ is inclination. 

The absolute Jacobian determinant of this transformation is
\begin{equation}
    \left\lvert \mathrm{det} \left( \frac{\partial(A, B, F, G)}{\partial(a_0, \omega, \Omega, i)} \right) \right\rvert = a_0^3\sin^3i
\end{equation}
For the injection set, we assumed isotropic inclination (Section \ref{ssec:inj_dist}), meaning its probability density is proportional to $\sin i$. This leaves us with a factor of $a_0^3\sin^2 i$ in the Jacobian. 

By Kepler's third law, $a_0$ is related to $M_{\rm WD}$ by 
\begin{equation}
    a_0 = P^{2/3} M_{\rm WD} (M_1 + M_{\rm WD})^{-{2/3}}
\end{equation}
up to a multiplicative constant. Taking the partial derivative with respect to $M_{\rm WD}$, we get an additional factor:
\begin{equation}
    \left(\frac{\partial a_0}{\partial M_{\rm WD}} \right)_{P, M_1} = a_0 \left( \frac{3 M_1 + M_{\rm WD}}{3M_{\rm WD} (M_1 + M_{\rm WD})} \right)
\end{equation}

Lastly, we assumed that the orbital phase, $\phi$, was uniform between $0$ and $2\pi$ (i.e., the probability density is a constant). Meanwhile, the Gaia samples are flat in periastron time, $T_p$. 
At a fixed period, 
\begin{equation}
    T_P = \frac{P_{\rm orb}}{2\pi} \phi
\end{equation}
such that
\begin{equation}
    \left\lvert \frac{\partial T_p}{\partial \phi}\right \rvert = \frac{P_{\rm orb}}{2\pi}
\end{equation} 
This introduces a $P_{\rm orb}$ factor in the full Jacobian. 

Putting all of this together, we get 
\begin{equation}
    \log J  = 4 \log a_0 + \log (3 M_1 + M_{\rm WD}) -\log M_{\rm WD} - \log(M_1 + M_{\rm WD}) + \log(1-\cos^2 i) + \log(P_{\rm orb})
\end{equation}
which we subtract from $p(\theta\mid\Lambda)$ for each system in the Monte Carlo sum to form the importance weights. 

\section{Varying Pr(red) boundary} \label{appendix:prredmax_test}

\begin{figure}
    \centering
    \includegraphics[width=0.98\linewidth]{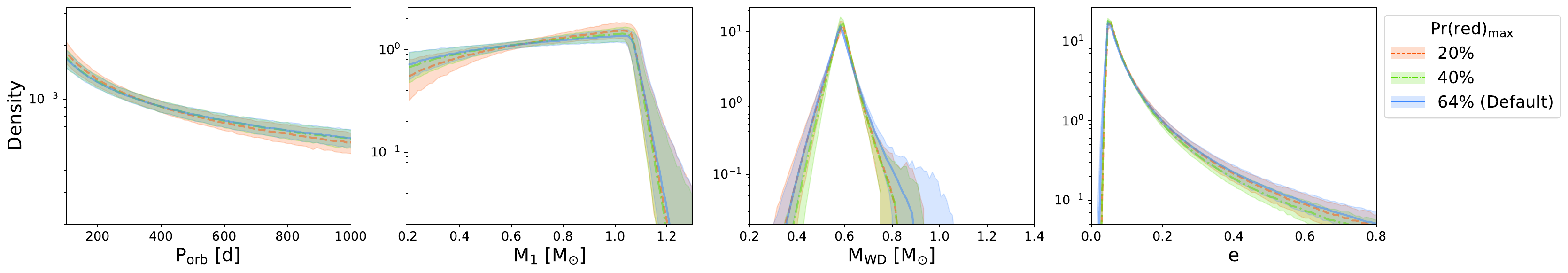}
    \caption{Comparison between the intrinsic parameter distributions resulting from the fit to observed samples assuming different values of $\rm Pr(red)_{max}$. Overall, there is little qualitative differences between the distributions, with the exception of the slope below the peak in $M_1$. }
    \label{fig:prred_1d_ppc}
\end{figure}

To construct the final (NCE) sample, \citetalias{Shahaf2024MNRAS} set an upper bound on $\rm Pr(red)$ of the NCI systems, $\rm Pr(red)_{max} = 64\%$, removing those that appear significantly redder than expected for a single MS star contributing to the observed flux. This serves to exclude triples with inner MSMS binaries which are expected to be the primary contaminant in the NCI sample (Section \ref{ssec:color_excess}). Here, we vary $\rm Pr(red)_{max}$ used in constructing the observed sample (Section \ref{ssec:color_excess}) and correspondingly in the detection criterion for the injection set (Section \ref{ssec:selection_func}), which then enter our hierarchical Bayesian model. In Figure \ref{fig:prred_1d_ppc}, we plot the 1D intrinsic parameter distributions resulting from these runs. 

The orbital period and eccentricity distributions are minimally sensitive to $\rm Pr(red)_{max}$. While there is variation at the low-mass end of the WD mass distribution, this is primarily captured by the B-spline components, which have large uncertainties. There is also no clear monotonic trend with $\rm Pr(red)_{max}$. 

Perhaps the most interesting feature is the slope below the peak of the $M_1$ distribution, which becomes increasingly flat with larger values of $\rm Pr(red)_{max}$. For a given secondary mass, in the limit of a dark object (e.g. WD), the implied primary semi-major axis is maximized to the observed photocentric semi-major axis, which in turn minimizes the inferred primary mass. This means that contaminating triples hosting luminous inner binaries, wrongly assumed to be WDs, will have underestimated primary masses, all else equal. This is qualitatively consistent with there being increased contamination at higher $\rm Pr(red)_{max}$, corresponding to a weaker cut on CE, which leads to there being more weight placed at lower $M_1$'s. Moreover, the best-fit inferred rate increases with $\rm Pr(red)_{max}$, also consistent with increased contamination implying a larger intrinsic number of systems. 

However, since all parameters from the three runs are still consistent with each other within $2\sigma$, we conclude that the differences are not statistically significant and will thus result in minimal changes to our conclusions. We leave further exploration of this effect to future work.

\section{Varying the age distribution} \label{appendix:triang_age}

\begin{figure}
    \centering
    \includegraphics[width=0.98\linewidth]{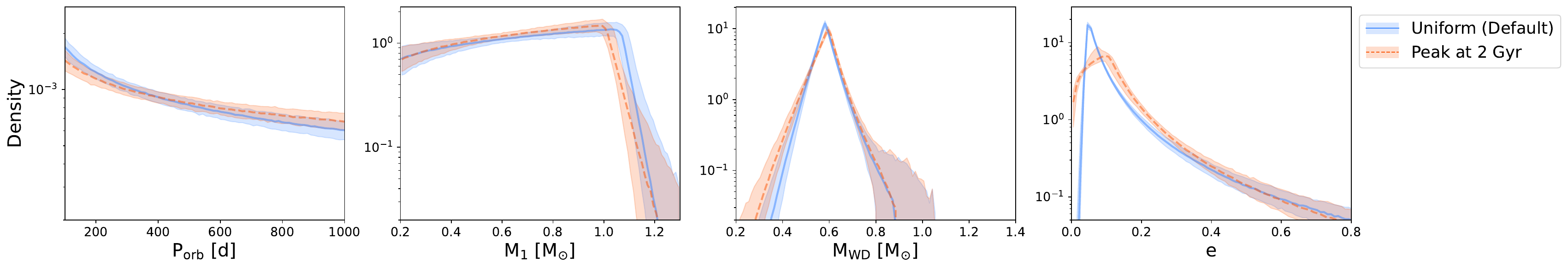}
    \caption{Analogous to Figure \ref{fig:prred_1d_ppc}, but comparing results assuming two different underlying age distributions.}
    \label{fig:age_dist}
\end{figure}

Instead of assuming a linearly uniform distribution in age (Section \ref{ssec:inj_dist}), we take a triangular distribution in logarithmic space peaking at $2\,$Gyr. This choice is motivated by evidence of a star formation burst that occurred $\sim 2-3\,$Gyr ago \citep{Mor2019A&A}, which also motivated the fixed $2\,$Gyr choice made by \citetalias{Shahaf2024MNRAS} to calculate predicted colors. In Figure \ref{fig:age_dist}, we plot intrinsic distributions derived using this injection set. 

Compared with results under the fiducial uniform age assumption, the new triangular age assumption yields a lower inferred space density ($2.6\times10^4$ vs. $3.5\times 10^4$) but similar intrinsic parameter distributions overall. The space density can be explained by the higher average detectability in the injection set ($\sim 17\%$) compared to the fiducial model ($\sim 13\%$). This is attributed to the color evolution of stars on the MS, as the triangular age distribution means more stars are younger than the $2\,$Gyr reference age used to evaluate the expected color. The same effect is likely responsible for the slight shifts in the peak of the $M_1$ (and subsequently, $M_{\rm WD}$) distribution. 

Interestingly, the biggest difference is in the low-eccentricity tails of the eccentricity distributions, where the new model is broader and peaked at a higher value. The origin of this difference is uncertain, but may reflect the fact that \texttt{gaiamock} implements an age-dependent function for the stellar luminosity which affects the astrometric precision. Therefore, the age likely affects the size of uncertainties on the derived orbital parameters, and thus the probability that a system passes the Gaia quality cuts to receive an astrometric orbital solution. This sensitivity could be enhanced for nearly circular orbits, where the degeneracy between the argument and epoch of periastron may affect the individual Thiele–Innes parameter uncertainties \citep{El-Badry2024OJAp}. A detailed investigation is beyond the scope of this work; the differences should be regarded as a systematic uncertainty associated with the assumed age distribution.

\section{Simple parametric model} \label{appendix:parametric}

\begin{figure*}
    \centering
    \includegraphics[width=0.98\linewidth]{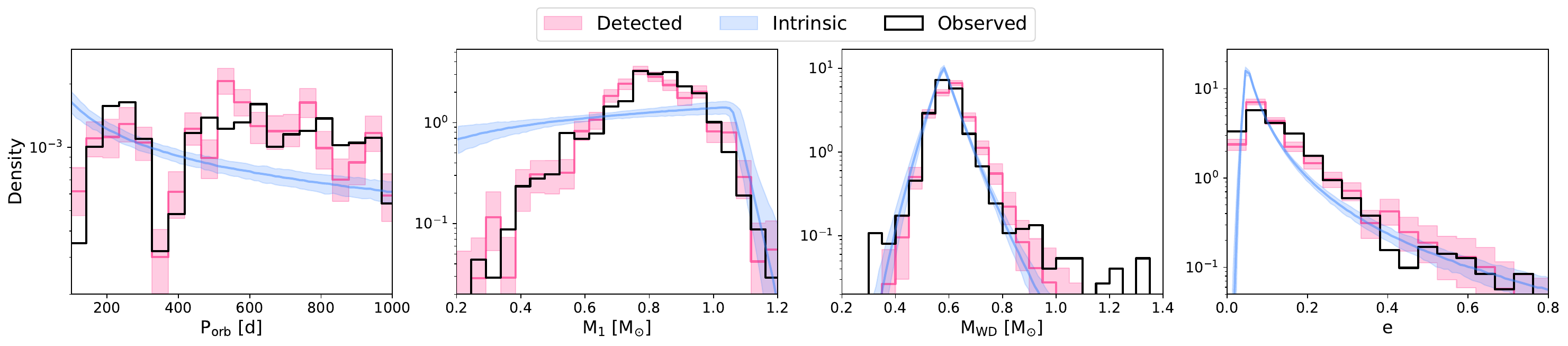}
    \caption{Analogous to plots in the upper row of Figure \ref{fig:ppc}, but without the B-spline component for $M_{\rm WD}$.}
    \label{fig:1d_ppc_no_bspline}
\end{figure*}

Here, we report the results of fitting the $M_{\rm WD}$ distribution with just the single-break power law model, without the B-spline component (Section \ref{ssec:pop_model}). In Figure \ref{fig:1d_ppc_no_bspline}, we plot the 1D distributions of the fitted parameters from the PPC.

As expected, the predicted detected sample is in close agreement with the observations near the peak but under-predicts the contribution at both extremes of the distribution. This is the motivation for adding a B-spline component to characterize the full range of WD masses. The best-fit values of the other parameters do not change significantly.   

\section{Corner plot} \label{appendix:corner}

\begin{figure}
    \centering
    \includegraphics[width=0.9\linewidth]{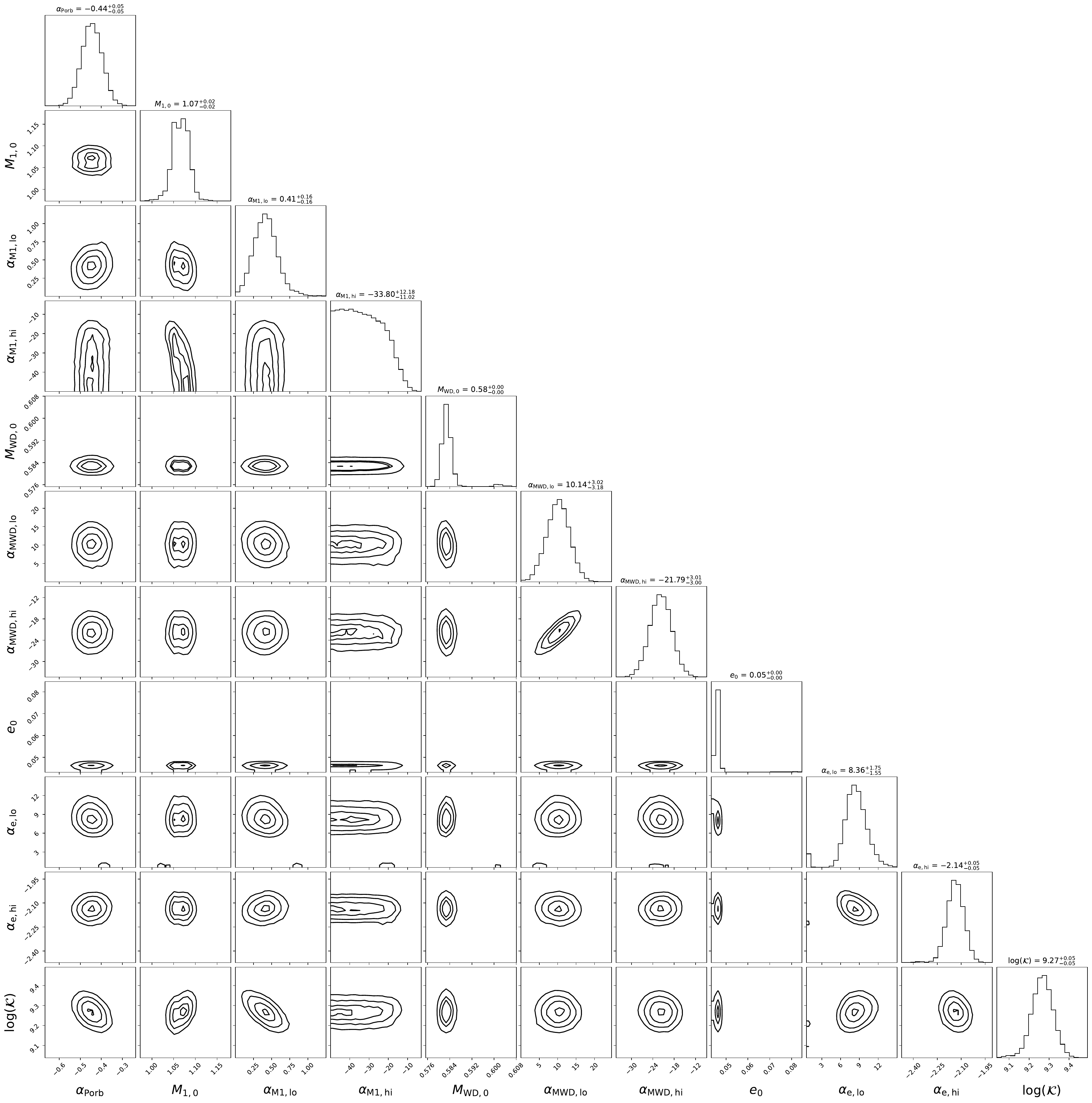}
    \caption{Corner plot showing results of MCMC fit for the fiducial model.}
    \label{fig:corner}
\end{figure}

In Figure \ref{fig:corner}, we show a corner plot of our fiducial results for parameters of the parametric backbone (i.e., spline coefficients omitted). Visually, the sampler is well-behaved. Marginal distributions are smooth and unimodal, and most off-diagonal contours are well-defined, with circular or ellipsoidal shapes. The slope of $M_1$ above the peak is poorly constrained, as there are few observed systems with $M_1 \gtrsim 1\,M_{\odot}$. 

The small secondary peaks in $M_{\rm WD, 0}$, $e_0$, and $\alpha_{e, \rm lo}$ correspond to a joint secondary region occupied by a minority of MCMC walkers. No transitions between the two regions were observed during production, so their relative posterior weights are not reliably determined. These structures may reflect alternative combinations of the break locations, slopes, and spline coefficients, although their origins have not been established.

\section{Testing different $M_1$ population models} \label{appendix:m1_models}

\begin{figure}
    \centering
    \includegraphics[width=0.95\linewidth]{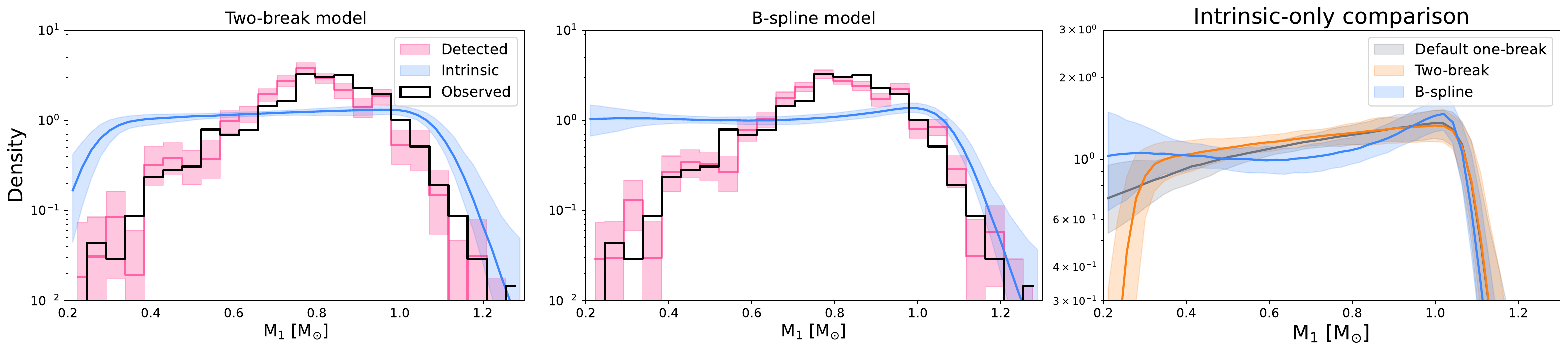}
    \caption{The first two panels are analogous to the $M_1$ plot in the upper row of Figure \ref{fig:ppc}, but fitting a two-break power-law (\textit{Left}) and a single-break power-law + B-spline (\textit{Center}) for the $M_1$ distribution. The rightmost panel compares the intrinsic distributions of these two models with the default single-break power-law model. All three infer a very shallow decline below a peak at $\sim 1\,M_{\odot}$.}
    \label{fig:placeholder}
\end{figure}

In the fiducial model, we fit a single-break power law to the $M_1$ distribution (Section \ref{ssec:pop_model}). This yields a shallow decline below the peak, which is in contrast to the findings of \citetalias{Yamaguchi2026arXiv} and is unexpected given a fixed critical mass ratio (Section \ref{ssec:y26_comparison}). Here, we test the robustness of this result by testing two additional models for $M_1$, which allow the slope to evolve below the peak: (1) a two-break power law, and (2) a single-break power law but with an added B-spline component. The results are shown in Figure \ref{fig:placeholder}. All three models imply a relatively flat distribution below the peak at $\sim 1.0\,M_{\odot}$, all the way down to $\sim 0.3\,M_{\odot}$. Therefore, we conclude that this feature of the $M_1$ distribution is not solely due to our modeling choice.  

We note that the two-break model leads to an additional sharp turnover at $M_1\sim0.27\,M_\odot$. However, this feature lies in a regime of low completeness containing only three observed systems below $0.3\,M_\odot$, and its steepness is weakly constrained. In particular, the broad positive-slope prior permits strong suppression below the lower break and may influence the inferred decline. Since this feature is not recovered by the spline model, we regard it as model-dependent rather than evidence for a physical cutoff.

\bibliographystyle{aasjournal}



\end{document}